\documentclass[preprint]{JHEP} 
\usepackage{epsfig}
\usepackage{amsmath}
\usepackage{amssymb,amsfonts}
\title{Heterotic/Type II Triality and \\Instantons on $K_3$}

\author{ Elias Kiritsis\\
Physics Department, University of Crete,\\ 71003 Heraklion,
GREECE\\ E-mail: \email{kiritsis@physics.uoc.gr}}
\author{Niels Obers\thanks{
Work supported in part by TMR network ERBFMRXCT96-0045.}\\ Nordita
and Niels Bohr Institute\\ Blegdamsvej 17, DK-2100 Copenhagen,
DENMARK\\ E-mail: \email{obers@nordita.dk}}
\author{Boris Pioline\footnote{On
leave of absence from LPTHE, Universit{\'e} Pierre et Marie Curie,
PARIS VI and Universit{\'e} Denis Diderot, PARIS VII, Bo\^{\i}te
126, Tour 16, 1$^{\it er}$ {\'e}tage, 4 place Jussieu, F-75252
Paris CEDEX 05, FRANCE}\\ Jefferson Physical Laboratory, Harvard
University\\ Cambridge, MA 02138, USA\\ E-mail:
\email{pioline@physics.harvard.edu}}

\preprint{\hepth{0001083}\\ NBI-HE-99-53 \\NORDITA-1999/79 HE \\
 HUTP-99/A069 \\LPTHE-99-49  }      

\def\be{\begin{equation}}
\def\ee{\end{equation}}
\def\bs{\begin{subequations}}
\def\es{\end{subequations}}

\newcommand{\nn}{\nonumber}

\def\d{{\rm d}}

\def\sp{\;\;\;,\;\;\;}

\def\e{\epsilon}

\def\et{\hat E_2}
\def\hd{\hat {\rm d}}
\def\ad{\dot\alpha}
\def\a{\alpha}
\def\b{\beta}
\def\c{\gamma}
\def\bd{\dot\beta}

\def\dd{\dot\delta}

\def\m\mu
\def\n{\nu}

\def\a{\alpha}

\def\pa{\partial}
\newcommand{\orb}{+\ \mbox{orb.}}
\newcommand{\eis}[3]{~\ensuremath{{\cal E}^{#1}_{\irrep{#2};#3}}}
\newcommand{\thet}[2]{\!\theta\left[ {}^{#1}_{#2} \right]}
\newcommand{\so}[2]{~\ensuremath{[SO(#1)\times SO(#2)] \backslash SO(#1,#2,\Real)}}
\renewcommand{\th}{\vartheta}
\newcommand{\ar}[2]{\left[ {}^{#1}_{#2} \right]}
\newcommand{\Zar}[2]{Z\!\left[ {}^{#1}_{#2} \right]}

\newcommand{\irrep}[1]{\ensuremath{\boldsymbol{#1}}}

\newcommand{\Tr}{{\rm Tr}}

\newcommand{\B}{{\cal B}}

\newcommand{\M}{{\cal M}}
\newcommand{\R}{{\cal R}}

\newcommand{\F}{{\cal F}}

\newcommand{\Zint}{\mathbb{Z}}
\newcommand{\Real}{\mathbb{R}}

\def\a{\alpha}
\def\n{\nu}

\def\pa{\partial}
\newcommand{\iia}{{\rm IIA}}
\newcommand{\iib}{{\rm IIB}}

\newcommand{\lh}{\ensuremath{l_{\rm H}}}
\newcommand{\rh}{\ensuremath{R_{\rm H}}}
\newcommand{\lm}{\ensuremath{l_{\rm M}}}
\newcommand{\lii}{\ensuremath{l_{\rm II}}}

\newcommand{\lp}{\ensuremath{l_{\rm P}}}
\newcommand{\gh}{\ensuremath{g_{\rm H}}}
\newcommand{\gii}{\ensuremath{g_{\rm II}}}

\newcommand{\giis}{\ensuremath{g_{\rm 6IIA}}}
\newcommand{\ghs}{\ensuremath{g_{\rm 6H}}}

\newcommand{\giibs}{\ensuremath{g_{\rm 6IIB}}}

\renewcommand{\sp}{\ ,\qquad}

\abstract{A detailed understanding of instanton effects for
half-BPS couplings is pursued in theories with 16 supersymmetries.
In particular, we investigate the duality between heterotic string
on $T^4$ and type IIA on $K_3$ at the $T^4/\Zint_2$ orbifold
point, as well as their higher and lower dimensional versions. We
present a remarkably clean quantitative test of the duality at the
level of $F^4$ couplings, by completely matching a purely one-loop
heterotic amplitude to a purely tree-level type II result. The
triality of $SO(4,4)$ and several other miracles are shown to be
crucial for the duality to hold. Exact non-perturbative new
results for type I', F on $K_3$, M on $K_3$, and IIB on $K_3$ are
found, and the general form of D-instanton contributions in type
IIA or B on $T^4/\Zint_2$ is obtained. We also analyze the
NS5-brane contributions in type II on $K_3\times T^2$, and predict
the value $\mu (N)=\sum_{d|N} (1/d^3)$ for the bulk
contribution to the index of the NS5-brane world-volume theory on
$K_3 \times T^2$.}


\begin{document}

\maketitle 

\section{Introduction}

\subsection{Instanton effects and BPS saturated couplings}
Understanding the rules of instanton calculus in string theory has
been a challenging goal over the last few years, the achievement
of which has become conceivable thanks to the D-brane description
of string solitons \cite{Polchinski:1995mt}. Yet, it is still a
difficult problem to compute the contribution of D-instantons to a
generic amplitude, not to mention that of NS5-brane instantons
relevant for four-dimensional physics. The study of exact BPS
saturated amplitudes ($R^4$ couplings being the primary example)
in a weak coupling expansion has shed a welcome light on this
problem, and indeed has allowed the first experimental
determination of the half-BPS D-instanton measure in the type IIB
theory \cite{Green:1998tn}, before the latter was derived from
first principles \cite{Moore:1998et,Krauth:1998xh}, together with
the leading perturbative corrections in the instanton background
\cite{Gava:1999zk}. U-duality \cite{Hull:1995ys} (see
\cite{Obers:1998fb} for a review) has been a prominent tool in
generalizing these results to toroidal compactifications of
M-theory \cite{Green:1997di,Pioline:1997pu,Obers:1999um}, and
there is by now a fairly complete understanding of half-BPS
instanton effects in these theories
\cite{Kiritsis:1999ss,Obers:1999um}, modulo still mysterious
effects superficially of order $e^{-1/g_s^2}$
\cite{Pioline:1997pu,Obers:1999um}.

The situation in theories with lower supersymmetry is however not
so well understood, even in the case of half-BPS amplitudes in
theories with 16 supersymmetries. The canonical examples in that
case correspond to $R^2$ couplings, believed to be one-loop exact
on the type II side, and $F^4$ couplings, believed to be one-loop
exact on the heterotic side
\cite{Yasuda:1988bu,Tseytlin:1996cg,Bachas:1997mc}. These
non-renormalization conjectures are supported by anomaly
cancellation arguments and decoupling between the gravitational
and vector multiplets. As far as $R^2$ couplings on the heterotic
side are concerned, the only half-BPS instanton is the heterotic
5-brane, and the lack of knowledge of its worldvolume dynamics has
hindered a direct understanding of its non-perturbative effects on
the heterotic side \cite{Harvey:1996ir,Gregori:1997hi}, even
though interesting results have been obtained on the type I side
\cite{Hammou:1999in}. We will instead focus on the $F^4$
couplings, for which several results are already available. On the
type I side, there is a quite complete treatment of the D-string
instanton contributions
\cite{Bachas:1997mc,Kiritsis:1997hf,Foerger:1998kw}, even though
some ill-understood higher genus contact contributions are needed
for the duality to hold \cite{Bachas:1997mc}. The $F^4$ couplings
on the type I' side have also been computed
\cite{Gutperle:1999dx,Gutperle:1999xu,Gava:1999ky}, but the
detailed instanton measure remains to be understood. They have
also been reproduced from the point of view of F-theory
compactified on $K_3$ at particular singular points of the moduli
space \cite{Lerche:1998nx,Lerche:1998gz,Lerche:1999hg}, but due to
the fact that the dilaton is fixed at a finite value, these
results give little insight into instanton effects. Finally,
closely related four-derivative scalar couplings in the context of
type II string theory compactified on $K_3$ have been obtained
\cite{Antoniadis:1997zt}, which are believed to be related by
supersymmetry to $F^4$ couplings. In the latter case, instanton
effects from D-branes wrapped on even homology cycles of $K_3$
have been identified, and shown to reproduce the type IIB
D-instanton contributions in the ten-dimensional
decompactification limit. The summation measure was recovered from
a D-brane matrix model in \cite{Green:1998tn}. NS5-brane
instantons were also found but not thoroughly discussed. It is the
purpose of this work to extend these partial results, solve
several of the issues raised above, and to try and achieve the
same level of understanding as in the maximally supersymmetric
case.

\subsection{Instantons on $K_3$ at the orbifold point}
In general, a detailed perturbative or instanton computation on a
curved manifold like $K_3$ is hampered by our lack of knowledge of
the $K_3$ stringy geometry beyond simple topological invariants.
Our main goal  is to obtain a working understanding of D-instanton
effects in type II theories compactified on $K_3$ at the
$T^4/\Zint_2$ orbifold point of $K_3$, for which the conformal
field theory description is completely solvable but still
non-trivial. $\Zint_{3,4,6}$ orbifold points are technically more
involved but expected to yield similar results. Other solvable
descriptions include Gepner points, but those do not in general
possess a well defined classical geometry limit, and one must
resort to boundary CFT techniques in order to understand these
stringy geometries \cite{Ooguri:1996ck}. In the simple
$T^4/\Zint_2$ orbifold case however, the geometric interpretation
is clear, and such techniques can be dispensed with. Instantons
simply arise from branes wrapped on even cycles of $T^4$, or
collapsed at the 16 orbifold singularities. They first show up in
type IIB compactified on $K_3$, or in IIA compactified on
$K_3\times S_1$ where the extra $S_1$ allows the even D-branes to
wrap a Euclidean submanifold. Translating the one-loop heterotic
result under the duality map, we shall obtain the contributions of
these D-instantons to the half-BPS saturated $F^4$ amplitudes.

Our method will appear to be equally applicable in any space-time
dimension. By going to the appropriate dual description, we will
obtained a wealth of complimentary information that we regard as
equally interesting. In $D=6$, we will obtain one of the cleanest
tests of heterotic-type IIA duality to our knowledge, by
recovering the one-loop result from a type IIA tree-level
amplitude. This is arguably the first non-trivial quantitative
test of heterotic-type II duality, since all other (with the
possible exception of \cite{Antoniadis:1997zt}, which will be
recovered in this work) follow from supersymmetry alone. In $D=7$,
we shall obtain the M-theory four-gluon amplitude for $SU(2)$
gauge bosons located at the $A_1$ singularities of $K_3$. In
$D=8$, we shall recover the $F^4$ amplitude for $SO(8)$ gauge
bosons located at the orientifold planes of Sen's F-theory model
\cite{Sen:1996vd}, and amend the existing knowledge
\cite{Lerche:1998nx}. In $D=9$, we will compute the $F^4$
couplings at the $SO(16)\times SO(16)$ point, and show that the
higher genus contact contributions found in \cite{Bachas:1997mc}
do not arise in this case. Finally, in $D=4$ we will obtain and
analyze the contribution of NS5-brane instanton effects, and
extract the corresponding instanton measure. We will also find an
interesting non-renormalization property beyond one-loop in the
background of the NS5-brane.

\subsection{A test of heterotic-type IIA duality}
For the convenience of the reader, we would like to sketch the
salient points of our analysis in the case of heterotic-type IIA
duality in six dimensions, which lies at the basis of our argument
and is quite representative of our method. We focus on $F^4$
couplings involving the 20 gauge fields from the vector
multiplets, disregarding the graviphotons for now, and more
specifically on the (0,16) of them originating from the Cartan
torus of the ten-dimensional gauge group. On the heterotic side,
$(\Tr F^2)^2$ couplings related by supersymmetry to Chern-Simons
couplings appear at tree-level already. We shall disregard them in
this work, since they are analogous to the $R^2$ couplings and
have a trivial dependence on the moduli. More interestingly, the
only further contributions to four-gauge-boson $F^4$ couplings
occur at one-loop on the heterotic side, and since they barely
saturate the fermionic zero-modes, they are given by the standard
integral on the fundamental domain $\F$ of the upper-half plane
\begin{equation}
\label{f4h} A^{\rm Het}_{F^4}= \lh^2 \int_{\cal F}
\frac{d^2\tau}{\tau_2^2} \frac{\bar Q^4\cdot
Z_{4,20}(g/\lh^2,b,y)}{\bar\eta^{24}} \ .
\end{equation}
Here $\lh$ is the heterotic string length, and is reinstated on
dimensional grounds. $Z_{4,20}$ denotes the partition function of
the heterotic even self-dual lattice of signature (4,20),
parameterized by the metric $g$ and Kalb-Ramond field $b$ on the
torus $T^4$ and the Wilson lines $y$ of the 16 $U(1)$ gauge fields
in ten dimensions along the 4 circles of the torus $T^4$. $\bar
Q^4$ denotes an operator inserting four powers of right-moving
momenta in the lattice partition function, depending on the 4
gauge fields considered, and $1/\eta^{24}=1/q+24+\dots$ is the
contribution of the 24 right-moving oscillators that generate the
Hagedorn density of half-BPS states in the perturbative spectrum
of the heterotic string.

Under duality with the type IIA theory compactified on $K_3$, the
six-dimensional string coupling $g_6$ gets inverted, while the
string length is rescaled as $l_s\to g_6 l_s$\footnote{In our
conventions, we transform the string length but leave the metric
invariant. This takes care of the Weyl rescalings needed to go
from the various string frames to the Einstein frame.}. Taking
into account the particular normalization of the type II Ramond
fields, it is easy to see that \eqref{f4h} translates into a
tree-level type IIA result. On the other hand, it is still given
by a modular integral on the fundamental domain of the upper-half
plane, which is usually characteristic of one-loop amplitudes. The
resolution of this paradox is that on the type IIA side, the gauge
fields dual to the $(0,16)$ heterotic ones originate from the
twisted sectors of the orbifold: the correlator of four $\Zint_2$
twist fields on the sphere can be re-expressed as the correlator
of single-valued fields on the double cover of the sphere, which
is a torus \cite{Dixon:1987qv,Hamidi:1987vh}; its modulus depends
on the relative position of the four vertices, and hence should
be integrated over. A careful computation yields the tree-level
type IIA result
\begin{equation}
\label{f4iia} A^{\iia}_{F^4}= \frac{1}{\gii^2} \gii^4
\frac{\lii^6}{V_{K_3}}
 \int_{\cal F} \frac{d^2\tau}{\tau_2^2} Z_{4,4}(G/\lii^2,B) \ ,
\end{equation}
where the factors of $\gii$ correspond to the tree-level weight
and the normalization of the Ramond fields respectively. Here we
have focused for simplicity on a particular choice of $(0,16)$
fields: in general, \eqref{f4iia} involves a shifted lattice sum
integrated on a six-fold cover ${\cal F}_2$ of the fundamental
domain ${\cal F}$.

Still this result is not quite of the same form as \eqref{f4h}.
For one thing, the type IIA result, being a half-BPS saturated
coupling, does not involve any oscillators, in contrast to the
heterotic side. For another, the \so{4}{4} moduli $G/\lii^2,B$ are
not the same as the heterotic $g/\lh^2,B$. In order to reconcile
the two, we need to take several steps:

\vskip 2mm \noindent {\it (i) Moduli identification: } The
relation between the heterotic and type II moduli can be obtained
by studying the BPS spectrum. On the heterotic side, the BPS
states are Kaluza-Klein and winding states transforming as a {\it
vector} of $SO(4,4,\Zint)$, and possibly charged under the 16
$U(1)$ gauge fields. On the type IIA side, a set of BPS states is
certainly given by the D0-, D2- and D4-branes wrapped on the even
cycles of $T^4$, which are invariant under the $\Zint_2$
involution. These states transform as a {\it conjugate spinor} of
the T-duality group $SO(4,4,\Zint)$, as D-branes should
\cite{Obers:1998fb}. We thus find that the heterotic $g/\lh^2,b$
and type IIA $G/\lii^2,B$ moduli should be related by $SO(4,4)$
{\it triality} \cite{Antoniadis:1999rm}, which exchanges the
vector and conjugate spinor representations.

There are also D2-brane states wrapped on the collapsed spheres at
the sixteen orbifold singularities \cite{Douglas:1996sw}, and
charged under the corresponding $U(1)$ fields. These are to be
identified with the charged BPS states on the heterotic side, and
their masses are matched by choosing the Wilson lines as
\cite{Bergman:1999kq}
\begin{equation}
\label{hety} y=\frac{1}{2}\begin{pmatrix} 0101 &0101 &0101 &0101\\
0000 &0000 &1111 &1111\\ 0000 &1111 &0000 &1111\\ 0011 &0011 &0011
&0011
\end{pmatrix} \ .
\end{equation}
This can also be derived by realizing that the Wilson lines along
the first circle in $T^4$ map to the B-field fluxes on the
collapsed two-spheres, which have been shown to be half a unit in
order for the conformal field theory to be non-singular
\cite{Aspinwall:1995fw}. If we instead put this Wilson line to
zero, we recover a gauge symmetry $SO(4)^8=SU(2)^{16}$, as
appropriate for the 16 $A_1$ singularities of $T^4/\Zint_2$. This
choice is relevant for M-theory compactified on $K_3$ at the
$\Zint_2$ orbifold point. If we further omit the Wilson lines on
the 2nd (resp 2nd and 3rd) circles, the gauge symmetry is enlarged
to $SO(8)^4$ (resp. $SO(16)^2$), which are relevant for F-theory on
$K_3$ and type I' respectively. These relations explain why our
results can easily been applied to these settings as well.

\vskip 2mm

\noindent{\it (ii) Hecke identities:} At the above choice of Wilson
lines, it so happens that the lattice sum simplifies drastically.
This phenomenon was noted in a particular example in
\cite{Lerche:1998nx}, and we will greatly extend its range of
validity. In order to see this, it is useful to reformulate the
above choice of Wilson lines on the heterotic $T^4$ as a
$(\Zint_2)^4$ freely acting orbifold, so that
\begin{equation}
\label{z420het}
Z_{4,20} = \frac{1}{2^4} \sum_{h,g} Z_{4,4}\ar{h}{g}
\bar\Theta_{16}\ar{h}{g}  \ ,
\end{equation}
where $g$ and $h$ run from 0 to 15 and are best seen as four-digit
binary numbers; $h$ labels the twisted sector while the summation
over $g$ implements the orbifold projection in that sector. The
blocks $Z_{4,4}\ar{h}{g}$ are partition functions of (4,4)
lattices with half-integer shifts, and $\bar\Theta_{16}\ar{h}{g}$ are
antiholomorphic conformal characters. The operator $\bar Q^4$ only
acts on the latter. As we shall prove in Appendix \ref{latsum},
extending techniques first developed in \cite{Mayr:1993mq}, the
conformal blocks $\Phi\ar{h}{g}=Q^4
\Theta_{16}\ar{h}{g}/\eta^{24}$ occurring in the modular integral
can be replaced by two-thirds their image $\lambda$ under the
Hecke operator
\begin{eqnarray}
H_{\Gamma_2^-}.\Phi(\tau)=\frac{1}{2} \left(
 \Phi\left(-\frac{1}{2\tau}\right) +
\Phi\left(\frac{\tau}{2}\right)+ \Phi\left(\frac{\tau+1}{2}\right)
\right)
\end{eqnarray}
provided this image is a constant real number:
\begin{equation}
\label{hecke} H_{\Gamma_2^-}\cdot \left[ \frac{Q^4
\Theta_{16}\ar{0}{1}}{\eta^{24}} \right]=\lambda \ .
\end{equation}
We observe that the relation \eqref{hecke} holds for all the
conformal blocks of interest in this construction. The modular
integral thus reduces to
\begin{equation}
\label{f4hr} A^{\rm Het}_{F^4}=  \frac{2\lambda}{3} \lh^2
\int_{\cal F} \frac{d^2\tau}{\tau_2^2}
 Z_{4,4}(g/\lh^2,b)
\end{equation}
and the Hagedorn density of half-BPS states in \eqref{z420het} has thus
cancelled. We note that modular integrals such as \eqref{f4hr}
have infrared divergences coming from the vacuum sector in the
lattice partition function, and we implicitly subtract the
divergent term. This is natural from the point of view of the
one-loop heterotic thresholds, and required from the point of view
of the tree-level type IIA result since we need to subtract the
tree-level exchange of massless modes to get the correction to the
two-derivative effective action.

\vskip 2mm

\noindent{\it (iii) Triality:} the last step needed to identify the type
IIA and heterotic result is to understand how triality equates the
integrals of the partition function $Z_{4,4}(g/\lh^2,b)$ and
$Z_{4,4}(G/\lii^2,B)$ on the fundamental domain of the upper-half
plane. It is easy to convince oneself that such an equality cannot
hold at the level of integrands, by looking at some
decompactification limits. However, it has been shown that such
modular integrals could be represented as Eisenstein series for
the T-duality group $SO(4,4,\Zint)$, in the {\it vector} or
(conjugate) {\it spinor} representations according to one's taste
\cite{Obers:1999um}:
\begin{equation}
\pi \int_{\cal F} \frac{d^2\tau}{\tau_2^2}Z_{4,4}
=\eis{SO(4,4,\Zint)}{V}{s=1}=\eis{SO(4,4,\Zint)}{S}{s=1}
=\eis{SO(4,4,\Zint)}{C}{s=1} \ .
\end{equation}
This implies the invariance of the modular integral of
$Z_{4,4}(g/l_s^2,b)$ under triality transformation of the moduli,
which completes the argument.

\subsection{Outline}
The previous discussion was intended as a preview only, and will
be made precise and generalized in the rest of the paper. The
latter is organized in such a way that the reader may skip the
more technical sections without major inconvenience. In Section 2,
we give an overview of the various dual descriptions of the
heterotic string compactified on a torus $T^d=T^{10-D}$ and derive
the precise duality maps involved. Section 3 will be devoted to a
derivation of the heterotic $F^4$ amplitudes and their cousins and
their representation in the freely acting orbifold language. In
Section 4 we will concentrate on the duality test sketched above,
and derive the type IIA tree-level amplitude. Section 5 will be
devoted to translating the one-loop heterotic results in $4\leq D
\leq 9$ to their respective dual descriptions, and to interpreting
these results as instanton effects. Useful facts involving modular
forms and shifted lattice partition functions are gathered in the
appendices, together with details on the computation of modular
integrals of shifted lattice partition functions.

\subsection{Note and Acknowledgements}
In the course of this project, we learnt that E. Gava, Narain K.
and C. Vafa had tackled this problem independently, and in
particular independently noticed that the type II tree-level
amplitude was a one-loop result in disguise; we are grateful to
Edi Gava for communicating some of their preliminary results. We
are also indebted to Stephan Stieberger for his assistance in
reconciling our approach with his results with W. Lerche
\cite{Lerche:1998nx}. We also learnt that W. Nahm and K. Wendland
independently found the triality of $SO(4,4)$ to be relevant for
describing the moduli space of $K_3$ at the $\Zint_2$ orbifold
point \cite{Nahm:1999ps}. We furthermore acknowledge helpful
discussions with F. Cachazo, M. Gutperle, W. Lerche and P. Mayr.

\section{Moduli identification\label{mod}}
In this section, we will discuss how the heterotic string theory
can be mapped to its various dual descriptions. We start by
briefly recalling some basic results about the heterotic moduli
space.

\subsection{Toroidal compactifications of the heterotic string\label{modh}}
We consider the $E_8\times E_8$ or $SO(32)$ heterotic string
theory compactified on a torus $T^{d}$. For $d\leq 5$, the moduli
space takes the form
\begin{equation}
\label{hetmod} \Real^+ \times \so{d}{d+16} /  SO(d,d+16,\Zint) \ ,
\end{equation}
where the first factor is parameterized by the T-duality invariant
dilaton $\phi_{10-d}$ related to the ten-dimensional heterotic
coupling $g_{\rm H}$ by $e^{-2\phi_{10-d}}=V_d / (g_{\rm H}^2
l_{\rm H}^d)$, with $V_d$ the volume of the $d$-torus; the second
factor is the standard Narain moduli space, describing the metric
$g$ and B-field $b$ of the internal torus, together with the
Wilson lines $y$ of the 16 U(1) gauge fields in the Cartan torus
of the ten-dimensional gauge group \cite{Narain:1986jj}. The right
action of the discrete group $SO(d,d+16,\Zint)$ (by which we mean
the automorphism group of the lattice $E_8\oplus E_8\oplus H^d$ or
$D_{16}\oplus H^d$ depending on the case, where $H$ is the
hyperbolic standard lattice) reflects the invariance under
T-duality. This moduli space is usually parameterized in the
Iwasawa gauge by the $SO(d,d+16,\Real)$ viel-bein
\begin{equation}
\label{hetbein} e_{\rm H}=\begin{pmatrix}v^{-t} & & \\ & 1_{16} & \\
& & v  \end{pmatrix} \cdot
\begin{pmatrix} 1_d & y & b-yy^t/2  \\ & 1_{16} & -y^t \\ & & 1_d
\end{pmatrix}\ ,\qquad
e_{\rm H}^t \eta e_{\rm H} =\eta\ ,\qquad \eta=\begin{pmatrix}
&&1_d\\&1_{16}&\\ 1_d&&\end{pmatrix}\ ,
\end{equation}
where $v$ is the viel-bein of the metric of the internal torus,
namely $g=\lh^2 v^t v $. Note in particular that $e_{\rm H}$
depends only on the dimensionless moduli $g/\lh^2,b$ and  $y$. The right
action by the $SO(d,d+k,\Zint)$ elements
\begin{equation}
\label{hetborel}
\begin{pmatrix} 1_d & y' & -y'y^{'t}/2 \\ & 1_{16} & -y^{'t} \\ & & 1_d
\end{pmatrix} \; \; \mbox{ and } \; \;
\begin{pmatrix} 1_d &   & b' \\ & 1_{16} &  \\ & & 1_d \end{pmatrix}
\end{equation}
preserves the Iwasawa gauge and generates the discrete Borel
symmetries
\begin{equation}
y\to y+y'\ ,\quad b\to b+ \frac{1}{2} ( y' y^t - y y^{'t})\quad
\mbox{ or } \quad b\to b+b' \ ,
\end{equation}
which should be supplemented by Weyl elements in order to generate
the full T-duality group.

In order to determine the mapping of moduli to the dual
descriptions, our main strategy will be to compare the BPS mass
formula on both sides. For perturbative heterotic BPS states, it
is simply given by
\begin{equation}
\label{masshet} {\cal M} ^2 = \frac{1}{\lh ^2} Q^t
(M_{d,d+16}-\eta) Q\ ,
\end{equation}
where $Q=(m^i,q^I,n_i)$ is the vector of momenta, charges and
windings and $M_{d,d+16}=e^t_{\rm H} e_{\rm H}$ in terms of the
viel-bein  \eqref{hetbein}. The charges $q^I$,$I=1\dots 16$ take
values in the even self dual lattice $E_8\oplus E_8$ or $D_{16}$.
The degeneracy $d(N)$ of states with $Q^t \eta Q=2m^i n_i +
(q^I)^2 =2N$ is given by the generating formula
\begin{equation}
\label{hetgen} \sum d(N) q^N = \frac{1}{\eta^{24}(\tau)} =
\frac{1}{q}+24 + \dots\ ,\qquad q=e^{2\pi i \tau}\ .
\end{equation}

This description of the moduli space is quite complete for
compactification down to 5 dimensions. For lower dimensional
compactification however, the moduli space increases due to the
dualization of the NS 2-form into a scalar $\theta$ (in four
dimensions), or of the 30 $U(1)$ gauge fields into scalars (in
three dimensions). As a result, the $\Real^+$ factor in
\eqref{hetmod} is enhanced to $U(1)\backslash Sl(2,\Real)$,
parameterized by a complex parameter $S=\theta+i/g_4^2$, acted upon
by $Sl(2,\Zint)$ S-duality transformations \cite{Font:1990gx},
whereas in $D=3$ all scalars are unified into a \so{8}{24}
symmetric manifold, acted upon by the U-duality group
$SO(8,24,\Zint)$ \cite{Sen:1995wr}. It would be quite interesting
to determine $SO(8,24,\Zint)$ invariant couplings in this case,
but we will not attempt to do this here. Instead, we will restrict
ourselves to $d\leq 6$, and focus on half-BPS saturated couplings
which depend on the heterotic Narain moduli only, and hence
receive contributions from one-loop only on the heterotic side.

As motivated in the introduction, we now would like to determine
the subspace of the moduli space \eqref{hetmod} dual to a
compactification on a flat space except for possible $\Zint_2$
conical singularities. It will turn out that such a description
exists only for particular values of the Wilson lines breaking the
$SO(32)$ gauge symmetry to a subgroup $SO(2^{5-p})^{p+1}$, $0\leq
p\leq 4$. Choosing $p$ Wilson lines out of the four lines
\eqref{hety} fulfills this condition, and so would of course any
permutation of the 16 vertical columns. For $d=4$, it would seem
that any generic value of $y$ breaking the gauge symmetry to
$U(1)^{16}$ would do, but this is not correct since $y$ should
respect a large discrete group of symmetries that we will discuss
in Section \ref{dual}. It would also seem that this same symmetry
breaking pattern (for $p>0$) could be obtained from  $E_8\times
E_8$ heterotic theory: however $E_8$ cannot be broken to $SO(16)$
by Higgs phenomenon but rather to $SO(14)\times U(1)$, and it is
necessary to go to an enhanced symmetry point to recover $SO(16)$
and its subgroups\footnote{We thank F. Cachazo for explaining this
to us.}. We therefore restrict ourselves to the more convenient
$SO(32)$ heterotic description. In the following we shall also
focus on the maximally broken situation $p=min(d,4)$, since other
cases, though interesting, can be obtained by straightforward
compactification and have been discussed in
\cite{Bachas:1997mc,Gutperle:1999dx}. Having fixed the values of
$y$ modulo 2, we see that the T-duality group is reduced from
$O(d,d+16,\Zint)$ to $O(d,d,\Zint)$, or rather to a finite index
subgroup of it.

\subsection{Type I'\label{modi}}
Let us first consider the compactification of the heterotic string
on a single circle of radius $R_{\rm H}$ with the Wilson line
$y=(0000 0000 1111 1111)$ breaking the gauge symmetry to
$SO(16)\times SO(16)$. This theory admits a dual description as
type IIA on the orientifold $S^1/\Zint_2$ of a circle of radius
$R_{\rm A}$, also known as type I' or IA \cite{Polchinski:1996df}.
The gauge symmetry arises from two groups of eight D8-branes located at
each of the fixed points. The mapping between the radii and string
length can be most easily obtained by first dualizing the
heterotic string to type I on a circle, $(g_s,l_s,R)\to
(1/g_s,g_s^{1/2} l_s,R)$, and then T-dualizing to type I'. In this
way we get
\begin{equation}
\label{dualip} g_{\rm I'}=\frac{l_{\rm H}}{g_{\rm H}^{1/2} R_{\rm
H}}\ , \quad l_{\rm I'}=g_{\rm H}^{1/2} l_{\rm H}\ , \quad
\frac{R_{\rm I'}}{l_{\rm I'}}=g_{\rm H}^{1/2} \frac{l_{\rm
H}}{R_{\rm H}} \ ,
\end{equation}
where the quantities on the left-hand side refer to the type I'
theory and those on the right-hand side to the heterotic theory.
In particular, the heterotic nine-dimensional coupling $g_9=\gh
(\lh/R_{\rm H})^{1/2}$, parameterizing the $\Real^+$ factor in
\eqref{hetmod}, becomes $(R_{\rm I'}/l_{\rm I'})^{5/4} g_{\rm
I'}^{-3/4}$, so that the factorization of the moduli space does
not seem to have a very natural interpretation on the type I'
side. Similarly, the mapping of the heterotic Wilson lines is
quite involved, and the duality map \eqref{dualip} is only correct
at the $SO(16)\times SO(16)$ point, to which we shall restrict
ourselves. The more general case is discussed in
\cite{Cachazo:2000ey}, where it is shown that a  real version of
$K_3$ underlies the type I' description. The $SO(16)\times SO(16)$
point should then correspond to the $T^4/\Zint_2$ orbifold point
of $K_3$. 

\subsection{F-theory on $K_3$ \label{modf}}
We now consider the $SO(32)$ heterotic string compactified on a
two-torus of K{\"a}hler class $T_{\rm H}=b+iV_{\rm H}$ and complex
structure $U_{\rm H}$, at the $SO(8)^4$ point, corresponding to a
choice of two Wilson lines in \eqref{hety}. Following the same
reasoning as above, we first dualize to type I on $T^2$ and then
apply a double T-duality on $T^2$ to go to type IIB on a
$T^2/\Zint_2$ orientifold, with moduli
\begin{equation}
g_{\rm B}=\frac{l_{\rm H}^2}{V_{\rm H}}\ ,\quad l_{\rm B}=g_{\rm
H} l_{\rm H}^2\ ,\quad V_{\rm B}=\frac{g_{\rm H}^2 l_{\rm
H}^4}{V_{\rm H}}
\end{equation}
and the same complex structure $U_{\rm B}=U_{\rm H}$. This is
precisely Sen's construction \cite{Sen:1996vd} of F-theory on
$K_3$ \cite{Vafa:1996xn} at the orbifold point $T^4/\Zint_2$, seen
as an elliptic fibration over the base $T^2/\Zint_2$ with a fiber
of complex modulus $U_{\rm F}=a+i/g_{\rm B}=T_{\rm H}$. The real
factor in \eqref{hetmod}, corresponding to the heterotic 8D
coupling, now parameterizes the size of the base $V_{\rm B}/l_{\rm
P}^2$ in 10D Planck units ($l_{\rm P}=g_{\rm B}^{1/4}l_{B}$),
while the \so{2}{18} moduli parameterize the complex structure of
elliptically fibered $K_3$'s. At the $T^4/\Zint_2$ orbifold fixed
point, an \so{2}{2} subspace remains available corresponding to
the $U_{\rm B}$ and $U_{\rm F}$ moduli, while the remaining
$2\times 16$ parameters are fixed at the value of the heterotic
Wilson lines. There exists other components in the F-theory moduli
space corresponding to a fixed dilaton $U_{\rm F}$ and describing
the other $\Zint_3,\Zint_4,\Zint_6$ orbifold points of $K_3$
\cite{Dasgupta:1996ij,Lerche:1998gz}, but we shall not describe
them here. We simply note that they give rise to exceptional gauge
symmetry, and are therefore better accommodated in the heterotic
$E_8\times E_8$ setting; the mapping is then obtained by a further
$(T_{\rm H},U_{\rm H})$ interchange on the heterotic side.

\subsection{Type IIA on $K_3$ \label{modii}}
The natural next step would be to discuss the dual of heterotic on
$T^3$, namely M-theory on $K_3$, but we shall find it more
convenient to consider heterotic on $T^4$ and its type IIA dual on
$K_3$ first, before taking the large coupling limit in the next
subsection.

We therefore consider the $SO(32)$ heterotic string compactified
on a torus $T^4$ with constant metric $g$ and B-field $b$, and for
now unspecified Wilson lines $y$. This theory is dual to type IIA
compactified on $ K_3$ \cite{Sen:1995cj} under the identifications
\begin{equation}
\lh=\giis \lii\ \ ,\quad \giis=\frac{1}{\ghs}\ ,\quad \left(
\frac{R_1}{l_{\rm H}} \right)^2 = \frac{V_{K_3}}{\lii ^4}\ ,
\label{hetiia}
\end{equation}
which can be obtained by identifying the IIA NS5-brane on $K_3$
with the fundamental heterotic string, and the type IIA D0-brane
with a heterotic Kaluza-Klein state along the circle of radius
$R_1$ in $T^4$. This requires breaking the $SO(4,20)$ symmetry to
$SO(3,19)$, and decomposing the viel-bein \eqref{hetbein} into
\begin{eqnarray}
\label{hetbein2} e_{\rm H} &=&
\begin{pmatrix}
\frac{\lh}{R_1} &  & & & \\ & v_3 ^{-t} & & & \\ & & 1_{16} & & \\
& & & v_3& \\ & & & & \frac{R_1}{\lh} \end{pmatrix} \cdot
\begin{pmatrix}
1 & A  & & & \\  & 1_3  & & & \\ & & 1_{16} & & \\ & & & 1_3 & -A^t
\\ & & & & 1 \end{pmatrix} \cdot \nonumber \\ &&\cdot
\begin{pmatrix}
1 & & & b_{13} & 0 \\ & 1_3  & & b_{33} & -b_{13}^t\\ & & 1_{16} &
& \\ & & & 1_3 &  \\ & & & & 1 \end{pmatrix} \cdot
\begin{pmatrix}
1 &0 & y_1 & -\frac{y_1 y_3^t}{2} & -\frac{y_1 y_1^t}{2} \\ & 1_3
& y_3 & -\frac{y_3 y_3^t}{2} & -\frac{y_3 y_1^t}{2}\\ & & 1_{16} &
-y_3^t& -y_1^t \\ & & & 1_3 & 0 \\ & & & & 1 \end{pmatrix} \ ,
\end{eqnarray}
where $v_3,b_{33}$ are the 3-torus viel-bein and B-field, $A$ and
$b_{31}$ are the off-diagonal metric $g^{1i} g_{11}$ and B-field
$b_{i1}$ ($i=2,3,4$); $y_1$  is the Wilson line around the first
circle and $y_3$ are the three Wilson lines around $T^3$.

On the type IIA side, the \so{4}{20} moduli space also has a
natural decomposition into $\Real^+ \times \so{3}{19}$, where the
first factor corresponds to the volume of $K_3$ in type IIA string
units, and the second parameterizes the unit volume Einstein
metric on $K_3$ (see \cite{Aspinwall:1996mn} for a review).
Together with the fluxes of the B-field along the 22 homology
2-cycles, these parameters make up an $SO(4,20)$ matrix
\begin{equation}
\label{iiabein} e_{\iia} =
\begin{pmatrix}\frac{\lii^2}{\sqrt{V_{K_3}}}& & \\ & e_{3,19} & \\ & &
\frac{\sqrt{V_{K_3}}}{\lii ^2}
\end{pmatrix} \cdot
\begin{pmatrix} 1 & B & -\frac{1}{2}B\eta_{3,19} B^t \\
& 1_{22} & -\eta_{3,19} B^t \\ & & 1 \end{pmatrix} \ ,
\end{equation}
where $e_{3,19}$ is the viel-bein parameterizing the Einstein metric of
$K_3$ and $\eta_{3,19}$
denotes the signature $(3,19)$ metric on the space of two-cycles
$H_2(K_3)$. This can also be obtained from the BPS mass formula
\cite{Ramgoolam:1998vc}
\begin{eqnarray}
\label{mass420} {\cal M} ^2 &=& \frac{1}{\giis^2 \lii ^2} q^t (e_{\rm IIA}^t
e_{\rm IIA}-\eta) q = \frac{1}{\gii ^2 \lii ^2} 
\left( q_0-\frac{V_{K_3}}{\lii ^4}
q_4 + B q_2 - \frac{B\eta_{3,19} B^t}{2} q_4\right)^2 \nn \\  &&+
\frac{V_{K_3}}{\gii ^2 \lii ^6} \left(q_2-\eta_{3,19} B^t
q_4\right)^t \left( e_{3,19}^t e_{3,19} -\eta_{3,19} \right)
\left(q_2-\eta_{3,19} B^t q_4\right) \ ,
\end{eqnarray}
where $q_0,q_2$ and $q_4$ denote the D0-,D2- and D4-brane charge.
Matching \eqref{iiabein} with \eqref{hetbein2} gives the
last identification in \eqref{hetiia}. The heterotic T-duality
group  $SO(4,20,\Zint)$ acting from the right on $e_{\iia}$ is now
interpreted as mirror symmetry of the $(4,4)$ $K_3$ superconformal
theory (see \cite{Nahm:1999ps} for a recent review), while the ADE
enhanced symmetries on the heterotic side arise from D2-branes
wrapped on vanishing cycles of $K_3$ on the type II side
\cite{Witten:1995ex}.

We now would like to identify the $T^4/\Zint_2$ orbifold point in
this moduli space. At that point, we have a very explicit
description of the 22 two-cycles in $H_2(K_3)$: 3 self-dual cycles
and 3 anti-self-dual cycles come from the 6 two-cycles in
$H_2(T^4)$, which are obviously invariant under the $\Zint_2$
involution which reverses the sign of the 4 coordinates, while 16
more anti-self-dual ones come from the collapsed two-spheres at
any of the 16 singularities. $H_2(T^4,\Zint)$ has a signature
(3,3) even inner product, given by the wedge product of two-forms
integrated on $T^4/\Zint_2$, and carries a natural metric
$M_{3,3}=G\wedge G/V_{K_3}$ orthogonal with respect to the inner
product (note that it is independent of the volume of $K_3$),
where $G$ is the metric on $T^4$. This
$SO(3,3)$ matrix is an alternative parameterization of the
unit-volume metric of $T^4$ perhaps less familiar than the
standard $G/(\det(G))^{1/4} \in Sl(4)$ representation, and is made
possible thanks to the isomorphism $SO(3,3)=Sl(4)$. In order to
match with the heterotic side, it is useful to rewrite it in the
standard form
\begin{subequations}
\label{bg}
\begin{gather}
M_{3,3}=\begin{pmatrix}\gamma ^{-1} & \gamma ^{-1} \beta \\ \beta
^t \gamma  ^{-1} & \gamma -\beta \gamma ^{-1}\beta
\end{pmatrix} \ ,\quad
\gamma=\frac{1}{ G^{11} V_{K_3}} \begin{pmatrix} G_{22} & G_{23} &
G_{24} \\ G_{23} & G_{33} & G_{34} \\ G_{24} & G_{34} & G_{44}
\end{pmatrix}\ ,
\\
\beta_{12}= G^{14}/G^{11}\ ,\quad \beta_{23}= G^{12}/G^{11}\
,\quad \beta_{31}= G^{13}/G^{11} \ ,
\end{gather}
\end{subequations}
where the matrix $M_{3,3}$ is written in the basis
$m^{34}$,$m^{42}$,$m^{23}$, $m^{12}$,$m^{13}$,$m^{14}$ of
$H_2(T^4)$, and $\gamma$ (resp. $\beta$)  are symmetric (resp.
antisymmetric) $3\times 3$ matrices. $G_{ij}$ is the metric on
$T^4$, and $G^{ij}$ the inverse metric. Decomposing the 3+16+3
B-fluxes into $B_3,B_{16},B_{\bar 3}$, we arrive at our final
parameterization of the moduli matrix at the $T^4/\Zint_2$ orbifold
point,
\begin{eqnarray}
\label{iiabeinf} e_{\iia}&=&\begin{pmatrix}
\frac{\lii^2}{\sqrt{V_{K_3}}} & & & & \\ & u ^{-t} & & & \\ & &
1_{16} & & \\ & & & u & \\ & & & & \frac{\sqrt{V_{K_3}}}{\lii^2}
\end{pmatrix} \cdot
\begin{pmatrix}
1 & B_3 & & & \\  & 1_3  & & & \\ & & 1_{16} & & \\ & & & 1_3 &
-B_3^t \\ & & & & 1 \end{pmatrix} \cdot \nonumber \\ &&\cdot
\begin{pmatrix}
1 & & & B_{\bar 3} & 0 \\ & 1_3  & & \beta & -B_{\bar 3}^t\\ & &
1_{16} & & \\ & & & 1_3 &  \\ & & & & 1 \end{pmatrix} \cdot
\begin{pmatrix}
1 &0 & B_{16} & -\frac{B_{16} \zeta^t}{2} & -\frac{B_{16}
B_{16}^t}{2} \\ & 1_3  & \zeta & -\frac{\zeta \zeta^t}{2} &
-\frac{\zeta B_{16}^t}{2}\\ & & 1_{16} & -\zeta^t& -B_{16}^t \\ &
& & 1_3 & 0 \\ & & & & 1 \end{pmatrix} \ ,
\end{eqnarray}
where $u$ is again a viel-bein for the metric $\gamma$, i.e. $u^t
u=\gamma$. It is now straightforward to identify the heterotic and
type IIA moduli \eqref{hetbein} and \eqref{iiabein}, and obtain
the complete duality map,
\begin{subequations} \label{trial1}
\begin{gather}
 V_{K_3}=R_1^2\ ,\quad \gamma=g_3\ ,\quad
\beta=b_{33}\ ,\quad B_3 = A \ ,\\ \quad B_{16}=y_1 \ ,\quad B_{\bar
3}=b_{13}\ ,\quad \zeta=y_3
\end{gather}
\end{subequations}
in respective string units. Forgetting for the moment the Wilson line
moduli, what we have obtained here is the triality mapping between
the $SO(4,4)$ matrices in the vector representation, as
appropriate for the heterotic side whose BPS states transform as a
vector of $SO(4,4)$, to the conjugate spinor representation, under
which the D-brane BPS states of type IIA on the untwisted cycles
of $T^4/\Zint_2$ transform. In order to appreciate this, it is
useful to consider rectangular tori with vanishing B-field, in
which case the mapping reduces to
\begin{eqnarray}
\begin{pmatrix} \ln R^{\rm H}_1 \\ \ln R^{\rm H}_2 \\ \ln R^{\rm H}_3 \\\ln R^{\rm H}_4 \end{pmatrix}
=P\cdot
\begin{pmatrix} \ln R^{\iia}_1 \\ \ln R^{\iia}_2 \\ \ln R^{\iia}_3 \\
\ln R^{\iia}_4 \end{pmatrix} \ ,\quad P=\frac{1}{2}
\begin{pmatrix} 1&1&1&1\\1&1&-1&-1\\1&-1&1&-1\\1&-1&-1&1\end{pmatrix}\ ,
\quad P^2=1\ ,
\end{eqnarray}
where $P$ acts as triality on the Cartan torus of $SO(4,4)$,
mapping the conjugate spinor $\irrep{C}$ to the vector
representation $\irrep{V}$. We also note that the triality acts on
the charges of these representations according to
\begin{equation}
\label{trial2}
(m_1;m_2,m_3,m_4;n^2,n^3,n^4;n^1)=(m;m^{34},m^{42},m^{23};
m^{12},m^{13},m^{14};m^{1234}) \ .
\end{equation}

We have however not entirely completed our duty, since we still
need to determine the values of the heterotic Wilson lines at the
$\Zint_2$ orbifold point. For this, we switch off the B-flux
$B_{16}$ on the 16 collapsed spheres as well as $B_3$. Due to the
16 $A_1$ singularities on the orbifold, the type IIA theory
develops an $SU(2)^{16}$ enhanced gauge theory. From
\eqref{trial1}, we see that this amounts to setting the heterotic
Wilson line $y_1$ to zero. The three remaining Wilson lines should
therefore be such that they break $SO(32)$ to $SU(2)^{16}$. This
is indeed the case for the Wilson lines in \eqref{hety}. The
choice of the fourth Wilson line may seem arbitrary, but this is
not so: on the type IIA side, the orbifold conformal field theory
$T^4/\Zint_2$ has a discrete symmetry group $G$ generated by the
$D_4$ dihedral discrete symmetry of the four $S^1/\Zint_2$ CFT's
\cite{Dijkgraaf:1988vp}, to be discussed further in Section
\ref{dualt}. As will become clear in Section \ref{hetaf1111},
\eqref{hety} is the only choice consistent with this symmetry. We
therefore deduce the corresponding values of the blow-up
parameters $\zeta$ and B-flux $B_{16}$ from \eqref{trial1}.

Let us briefly discuss the case of the $\Zint_3$ orbifold point of
$K_3$. In that case, there are 9 $A_2$ singularities, so the
symmetry group is enhanced to $SU(3)^{9}$ in the absence of
discrete B-flux. This has rank $18$, and can therefore only happen
at an enhanced symmetry point on the heterotic side. Moreover,
$SU(3)^9$ cannot be embedded in $SO(32)$. It can however be
embedded in $E_6\times E_6\times E_6$ (the exceptional gauge
symmetry found in \cite{Dasgupta:1996ij} for F-theory on $K_3$ at
the $\Zint_3$ orbifold point), which is an enhanced symmetry of
the $E_8\times E_8$ heterotic string. It is thus possible to
identify the $T^4/\Zint_3$ orbifold point in the $K_3$ moduli
space in an analogous way as we did, but we shall refrain from
attempting this here.

\subsection{M-theory on $K_3$ \label{modm}}
As announced, we now recover the dual description of the heterotic
string compactified on $T^3$ by decompactifying the heterotic
circle of radius $R_1$ in the above description. Since momentum
states along this circle are mapped to type IIA D0-branes, this is
the limit that takes type IIA on $K_3$ to M-theory on $K_3$, with
eleven-dimensional Planck length $l_{\rm M}^3=g_{\iia} \lii^3$
\cite{Witten:1995ex,Townsend:1995kk}. We therefore obtain
\begin{equation}
\label{dualm}
 l_{\rm M}^3= \frac{ g_{\rm H}^2 l_{\rm H}^6}{V_3}\ ,\qquad
V_{K_3}=\frac{g_{\rm H}^4 l_{\rm H}^{10}}{V_3^2} \ ,
\end{equation}
so that the fundamental heterotic string is identified with the
M5-brane wrapped on $K_3$. The $\Real^+$ factor in \eqref{hetmod}
now parameterizes the volume of $K_3$ in eleven-dimensional Planck
units,
\begin{equation}
 e^{4\phi_7/3}=V_{K_3}/\lm^4 \ ,
\end{equation}
while the \so{3}{19} moduli still describe the unit-volume
Einstein metric of $K_3$. At the $T^4/\Zint_2$ point, the same
parameterization as in \eqref{iiabein} is valid, restricted to the
$SO(3,19)$ subspace:
\begin{equation}
\label{mbein} e_{\rm M}=\begin{pmatrix} u ^{-t} & & \\ & 1_{16} &
\\ & & u  \end{pmatrix} \cdot
\begin{pmatrix}
1_3 & \zeta & \beta- \frac{\zeta \zeta^t}{2} \\ & 1_{16} &
-\zeta^t \\ & & 1_3
\end{pmatrix}\ ,
\end{equation}
so that the identification with the heterotic parameters is simply
\begin{gather}
\label{hetm} \gamma=g/\lh^2\ ,\quad \beta=b\ ,\quad \zeta=y \ .
\end{gather}
Whereas the mapping \eqref{trial1} could be seen as the statement
of triality, the identification \eqref{hetm} can be seen as the
realization of the exceptional isomorphism $SO(3,3)=Sl(4)$. Note
that all the B-field parameters have disappeared, in accordance
with the fact that M-theory does not possess any 2-form in its
spectrum, nor does $K_3$ have any three-cycle. In particular, this
implies that the 16 singularities are no more resolved by the
half-unit B-flux, and therefore a $SU(2)^{16}$ symmetry is
expected, arising from the M2-branes wrapped on the collapsed
spheres. This is the case if one  chooses the Wilson lines $y$ 
as the last three in \eqref{hety}.

\subsection{Type IIB on $K_3$ \label{modiib}}
We now turn to the five-dimensional compactification of the
heterotic string on $T^5=T^4\times S^1$, with the four Wilson
lines \eqref{hety} along $T^4$, breaking the gauge symmetry to
$U(1)^{16}$. This is dual to type IIA on $K_3\times S^1$ from
Section \ref{modii}, but we are interested here in the type IIB
description obtained by a further T-duality. Using the standard
$R\to l_s^2/R,g\to g l_s/R$ transformation rules, we find
\begin{equation}
\label{dualiib} \giibs=\frac{\lh}{R_{\rm H}}\ ,\quad \lii=\lh
\ghs\ ,\quad R_{\rm B}=\frac{\lh^2\ghs^2}{R_{\rm H}} \ ,
\end{equation}
so that in particular the heterotic five-dimensional dilaton is 
mapped to the size of the type IIB circle in 6D type IIB Planck
units,
\begin{equation}
g_{\rm H5} = R_{\rm B}/l_{\rm P}\ ,\quad l_{\rm P}^2=\giibs \lii
^2\ .
\end{equation}
The \so{5}{21} moduli, on the other hand, do not involve the
circle direction, and actually give the moduli space of the
six-dimensional type IIB theory compactified on $K_3$ only. The
full moduli space is obtained from the $\so{4}{20} $ moduli by
adjoining the six-dimensional type IIB coupling, together with the
fluxes ${\cal B}$ of the Ramond Ramond even forms on the 4+20
even-cycles of $K_3$:
\begin{equation}
\label{iibbein} e_{\iib} =
\begin{pmatrix} \giibs & & \\ & e_{4,20} & \\ & & \frac{1}{\giibs}
\end{pmatrix} \cdot
\begin{pmatrix} 1 & \B & -\frac{1}{2}\B\eta \B^t \\
& 1_{24} & -\eta \B^t \\ & & 1 \end{pmatrix} \ .
\end{equation}
The even forms wrapped on the same 4+20 cycles with two directions
less give 4+20 two-form gauge potentials $H$ with self-dual and
anti-self-dual field-strength respectively. The right-action by
$SO(5,21,\Zint)$ matrices corresponds to the U-duality symmetry of
type IIB on $K_3$. In Section \ref{instii}, by mapping  one-loop
$F^4$ couplings in five-dimensional heterotic string to type IIB
on $K_3\times S_1$ and taking the decompactification limit $R_{\rm
B}\to\infty$, we shall be able to derive the exact U-duality
invariant $t_{12} H^4$ couplings between 4 self-dual or
anti-self-dual two-forms in IIB compactified on $K_3$, and analyze
the resulting instanton contributions.

\subsection{Type IIA and IIB on $K_3\times T^2$ \label{mod4}}
Finally, we want to briefly discuss the duality between the
heterotic string on $T^6$ and type II theories on $K_3 \times
T_2$. This duality can be obtained straightforwardly by 
compactification from the previously discussed ones, and yields
the following identifications:
\begin{equation}
S_{\rm H}=T_{\iia}=U_{\iib}\ , \quad T_{\rm H}=S_{\iia}=S_{\iib}\
,\quad U_{\rm H}=U_{\iia}=T_{\iib}
\end{equation}
between the four-dimensional couplings $S=a+i/g_4^2$, K{\"a}hler
class $T$ and complex structure $U$ of $T^2$, with the string
scales related by $l_{\iia}=l_{\iib}=\lh \sqrt{T_{\rm 2H}/S_{\rm
2H}}$. $S_{\rm H}$ and its images parameterize the $U(1)\backslash
Sl(2,\Real)$ part of the moduli space, while $T_{\rm H}$ and
$U_{\rm H}$ arise in the decomposition of $SO(6,22)$ into
$SO(2,2)\times SO(4,20)$. This will enable us to obtain the
NS5-brane instanton contributions to $F^4$ couplings on the type
IIA and B side in Section \ref{inst4}, and in particular extract
the summation measure in \eqref{mu5}.

\section{Heterotic amplitudes \label{heta}}
Having identified the subspace of moduli space dual to $\Zint_2$
orbifold in various dimensions, we now would like
to compute the one-loop contribution on the heterotic side
for half-BPS saturated amplitudes, including the four-derivative couplings
\begin{equation}
\label{f4} t_8 \Tr F^4 \ ,\quad t_8 \Tr (F^2)^2\ ,
\end{equation}
where $F$ denotes the field strength of the $d+16$ right-moving
gauge bosons or the $d$ left-moving graviphotons,
as well as the couplings involving the gravitational sector,
\begin{equation}
\label{rr4} t_8 \Tr R^2 \Tr F^2 \ , \quad  t_8 (\Tr R^2)^2 \ , \quad
 t_8 \Tr R^4 \ ,
\end{equation}
where $t_8$ is the familiar eight-index tensor arising in various
string amplitudes \cite{Schwarz:1982jn}.

Before proceeding with the computation, it is probably worthwhile
recalling the arguments supporting the non-renormalization of
these couplings beyond one-loop on the heterotic side
\cite{Yasuda:1988bu,Tseytlin:1996cg,Bachas:1997mc}. First, 
in ten dimensions these terms are related by supersymmetry to CP-odd 
couplings such as $B\wedge \Tr F^4$, which should receive
no corrections beyond one-loop for anomaly cancellation.
A more explicit proof can be given at the level of string
amplitudes \cite{Yasuda:1988bu}, and goes through in lower
dimensions as well \cite{Bachas:1997mc}. 
This argument does not apply to the particular
combination $t_8 t_8 R^4=t_8(4\Tr R^4- (\Tr R^2)^2)$,
which forms a superinvariant on its own and could therefore receive
higher perturbative corrections.
Second, the only heterotic half-BPS instanton is the
heterotic 5-brane, which needs a six-cycle to wrap in order
to give a finite action instanton effect. For $d<6$ there can
therefore be no non-perturbative contributions beyond
the one-loop result. 
Third, it is
consistent with the factorization of the moduli space
\eqref{hetmod} and the T-duality symmetry $O(d,d+16,\Zint)$ to
assume that $t_8 \Tr F^4$ couplings are given at one-loop only and 
hence independent of the $\Real^+$ 
factor. In $d=6$ it is plausible
that supersymmetry prevents the mixing of the $Sl(2,\Real)$
dilaton factor with the Narain moduli in $F^4$ couplings, in the
same way as neutral hypermultiplets decouple from vector
multiplets in $N=2$ supergravity, and prevents corrections from
NS5-brane instantons \cite{Kiritsis:1999ss}. 
For $d=7$, U-duality mixes the dilaton with the Narain moduli, so 
that a similar statement cannot hold. Gauge fields being Poincar\'e dual
to scalars in 3 dimensions, the $F^4$ couplings translate into 
four-derivative scalar couplings, and should receive non-perturbative
corrections.
We will therefore assume that for all $d\leq 6$, 
the $F^4$ amplitudes involving four
right-moving gauge fields are given at one-loop only on the
heterotic side, and disregard a possible tree-level contribution for 
$t_8 \Tr (F^2)^2$ couplings.

Based on power counting, the $F^4$ couplings are clearly half-BPS
saturated, and the same will appear to be true for their $R^2 F^2$
and $R^4$ cousins. Indeed, from the point of view of the heterotic
world-sheet, space-time supersymmetry arises from the left-moving
sector, and gravitons are on the same footing as gauge bosons.
This is not so obvious on the type II side, where
part of the gauge bosons arise from the twisted Ramond-Ramond sector while the
gravitons come from the untwisted Neveu-Schwarz sector. It has
however been argued that $R^4$ and more generally 
$R^4 F^{4g-4}$ couplings were purely
topological for type IIA on $K_3$, and therefore should be half-BPS
saturated as well \cite{Berkovits:1998ex}.

For the uncompactified heterotic string, the couplings
\eqref{f4},\eqref{rr4} have been computed in
\cite{Lerche:1987sg,Lerche:1988zy,Ellis:1988dc} and shown to
involve the zero-modes of the right-moving currents only, reducing
to an elliptic genus. It is straightforward to adapt these
computations to toroidal compactifications, and in particular
to compactifications with discrete Wilson lines as in 
\eqref{hety}. This is what we now discuss, with a particular
emphasis on the miraculous simplifications that occur
and allow the heterotic-type II duality to hold.

\subsection{Orbifold partition function \label{hetao}}

In order to take advantage of the simple half-integer values
of the Wilson lines \eqref{hety}, we shall follow \cite{Gava:1999ky}
and describe the compactification on a torus $T^d(g,b)$
with $d$ Wilson lines from \eqref{hety}
as the $(\Zint_2)^d$ freely acting orbifold of a torus of double radius
by the $\Zint_2$ actions which combine a half-period
translation on each circle with the corresponding half-integer shift
on the lattice. This breaks the $SO(32)$ symmetry to $2^{d}$ copies
of $SO(2^{5-d})$, as we want. More explicitly, we decompose the partition
function of the $(d,d+16)$ lattice as
\begin{equation}
\label{orbpart} Z_{d,d+16}(g,b,y)=\frac{1}{2^{d}}
\prod_{i=1}^d \sum_{h^i,g^i=0}^1  Z_{d,d}\ar{h^1 \ldots h^d }{g^1
\ldots g^d}(g,b) \bar\Theta\ar{h^1 \ldots h^d}{g^1 \ldots
g^d}(0) \ .
\end{equation}
Here, $Z_{d,d}\ar{h}{g}$ is the $T^d$ lattice partition function,
with insertions of $(-)^{m_i g^i}$ and winding shifts $n^i\to
n^i+h^i/2$, while $\bar \Theta$ is given in terms of the usual
$\theta$-functions as
\begin{equation}
\label{shiftt} \bar \Theta\ar{h}{g}(\{v_I\})= \frac12 \sum_{a,b=0}^1
\prod_{\d=0}^{2^{d}-1} \left(\prod_{I=0}^{2^{4-d}}
 \thet{a+{\rm bin}(\d)\cdot h}{b+{\rm bin}(\d)\cdot g}(v_I^{\d}) \right) \ ,
\end{equation}
where ${\rm bin}(\d)$ is the $d$-digit binary representation of
$\d$. We have split the $2^5$ fermions representing the
$SO(32)$ current algebra into $2^{d}$ blocks of $2^{5-d}$ fermions
each. The arguments $v_I^{\d}$ allow to switch on a gauge
background $F_R^I=v_I$ in the $I$-th direction of the Cartan torus
of the $\d$-th copy of the gauge group $SO(2^{5-d})$, and will be
useful in deriving the elliptic genus shortly. In particular, for
$d=0$ we recover the $SO(32)$ lattice partition function. Setting
all the $v$'s to zero, the partition function reads
\begin{equation}
\label{zv0} Z_{d,d+16}(g,b,y)=\frac{1}{2^{d+1}} \bar
\theta_{\alpha}^{16} Z\ar{0}{0}+ \frac{1}{2^d} \left(
\bar\th_3^{8}\bar\th_4^{8} Z\ar{0}{\hd} +
\bar\th_3^{8}\bar\th_2^{8} Z\ar{\hd}{0} +
\bar\th_2^{8}\bar\th_4^{8} Z\ar{\hd}{\hd} \right) \ .
\end{equation}
Here, we have adopted a ``modular Einstein convention'' whereby
$\alpha=2,3,4$ is summed over all even spin structures and
$\d=0\dots 2^{d}-1$ is summed over all $d$-digit nonnegative
numbers, strictly positive if hatted. The three terms in the
parenthesis form an orbit of $Sl(2,\Zint)$, and we will henceforth
content ourselves with writing the first term $\orb$ only. We also
drop the $d,d$ subscript on $Z$ when no ambiguity is possible.
Note also that thanks to \eqref{s1}, the first unshifted term in \eqref{zv0}
can be distributed to the three shifted terms if we need to.

\subsection{Elliptic genus for higher derivative couplings \label{hetael}}
We now would like to adapt the computation of \cite{Ellis:1988dc}
to our particular toroidal compactification. Since the amplitude
is half-BPS saturated, the left-moving part of the four-gauge-boson (or
graviton) amplitude merely provides the kinematic structure, whereas
the right-moving currents reduce to their zero-mode part. Focusing
on the four-point amplitude for right-moving bosons first, we
therefore obtain
\begin{align}
\label{f40d} \int_{\cal F} \frac{d^2\tau}{\tau_2^2}& \sum_{p_L,p_R
\in \Gamma_{d,d+16}} t_8 F_R^I F_R^J F_R^K F_R^L \nn \\ &\left(
p_R^I p_R^J p_R^K p_R^L -\frac{6}{2\pi \tau_2} p_R^I p_R^J
g^{KL} +\frac{3}{4\pi \tau_2^2} g^{IJ} g^{KL} \right)
 ~\tau_2^{d/2} ~
\frac{q^{\frac{p_L^2}{2}} \bar q^{\frac{p_R^2}{2} }}
{\bar\eta^{24}} \ ,
\end{align}
where $F_R^I$ stands for the field-strength of any of the $d+16$
right-moving gauge bosons in the  Cartan torus of the gauge group.
Note in particular that this is expression is both modular
invariant and covariant under T-duality. The Dedekind function
$1/\bar \eta^{24}$ in \eqref{f40d} is the contribution of the 24
right-moving oscillators, which generate the tower of perturbative
half-BPS states.
The integral in \eqref{f40d} is actually infrared-divergent and should be
regularized. We assume in the following that this is done. 

In order to further simplify this expression, we must now
distinguish between the $(0,16)$ right-moving bosons coming from the
lattice $D_{16}$ and the $(0,d)$ from the torus. In the first case, the
insertion of a momentum $p_R^I$ amounts to taking a derivative
in \eqref{shiftt} with respect to the appropriate $v_I$. The non-holomorphic
contributions in \eqref{f40d} correct these derivatives
$\partial/\partial v$ into modular covariant derivatives
$\partial/\partial \hat v$. We can therefore omit them and
reinstate them at the end by covariance. In the case of the
$(0,d)$ gauge bosons, it is more convenient to perform a Poisson
resummation on the momenta in the lattice sum: the insertion of
$p_R^i$ then amounts to inserting $(m^i-\tau n^i)/\tau_2$, which
has modular weight $(0,1)$ as it should. Finally, in the case of
gravitons and graviphotons, the analogous statement is that one
should allow for a curvature background, thereby inserting
a factor $\xi(z)=z\eta^3 e^{-z^2/(8\pi \tau_2)}/\theta_1(z)$
for each pair of space-time coordinates,
and take derivatives with
respect to $z_i$ for each insertion of a gravi(pho)ton with
helicity in the $i$-th direction.  We will quote the
results in terms of the integrand $\Xi$ such that the
modular integral
\begin{equation}
\label{delta}
 \Delta_{F^4} = \frac{1}{{\cal{N}} \cdot 2^{d+1}} \int_{{\cal F}}
\frac{d^2\tau}{\tau_2^2}
 \frac{\bar\Xi_{F^4}}{\bar\eta^{24} }
\end{equation}
gives the higher-derivative coupling $t_8 \Tr F^4$ in the
effective action, where $F$ stands for either a gauge
field-strength or the Riemann tensor (seen as an $SO(10-d)$
field-strength), and the trace structure will be made precise. We
will denote by $F_\d$ the field-strength of the right-moving gauge
field in the $\d$-th copy of the gauge group, by $F_i$ the
right-moving $U(1)$ gauge fields from the torus, and by $G_i$ the
left-moving graviphotons.  The overall normalization will not be
fixed, but we will keep track of the relative normalization of the
various couplings, through the combinatorial factor ${\cal{N}}$.
We will then make use of the modular identities in Appendix
\ref{modu} as well as the theorem proven in Appendix \ref{latint}
to simplify these results and bring them in a form appropriate for
(i) comparison with the type II tree-level amplitude in  Section
\ref{dual} and (ii)  explicit evaluation and simplification for
the comparison with the other dual models in Section \ref{inst}.

\subsubsection{$\Tr F_{\d}^4$ and $(\Tr F_{\d}^2)^2$ \label{hetaf4}}
We start with the four-point amplitude of gauge bosons in a single
copy of $SO(2^{5-d})$. For $d<4$, this is a non-Abelian gauge group,
and we must therefore be careful with the identification of the
trace structure. Expanding \eqref{orbpart} to fourth order in the
$v_I^{\d}$ for fixed $\d$, the $\Tr F_{\d}^4$ combination corresponds
to $\sum_I (v_I^{\d})^4$, while $(\Tr F_{\d}^2)^2$ corresponds to
$(\sum_I (v_I^{\d})^2)^2$. We thus get
\begin{subequations}
\begin{equation}
\label{xi4}
\Xi_{\Tr F_{\d_1}^4}=\th_\alpha^{15} \th_{\alpha} ^{''''} \Zar{0}{0}+
\left[
 \left( \frac{\th_3^{''''}}{\th_3} +\frac{\th_4^{''''}}{\th_4}
\right) \th_3^8 \th_4^8 \Zar{0}{\hd} \orb \right] - 3 \Xi_{(\Tr
F_{\d_1}^2)^2} \ ,
\end{equation}
\begin{equation}
\label{xi22}
\Xi_{(\Tr F_{\d_1}^2)^2}=\th_\alpha^{14} (\th_{\alpha} ^{''})^2 \Zar{0}{0}+
\left[
 \left(
\left(\frac{\th_3^{''}}{\th_3}\right)^2
+\left(\frac{\th_4^{''}}{\th_4}\right)^2
\right) \th_3^8 \th_4^8 \Zar{0}{\hd} \orb \right] \ ,
\end{equation}
\end{subequations}
where $\orb$ denotes the two extra terms obtained from the first by
applying $S$ and $ST$ modular transformations.
For $d=4$ and higher, the gauge group is $U(1)$ and there is no
distinction between the two structures. Instead, the coupling
$(F_{\d_1})^4$ is given by $\Xi_{\Tr F_{\d_1}^4}  $ in \eqref{xi4}
without the $\Xi_{(\Tr F_{\d_1}^2)^2 }$ subtraction. We will discuss
in detail the procedure by which we simplify  these integrands in
the case of the first term in \eqref{xi4}, which we define as
$\Xi_{(F_{\d_1})^4}$. Other cases can be treated similarly and we
will only quote the final result.

Using the summation identity \eqref{s3}, we
can write the prefactor of $\Zar{0}{0}$ as $\eta^{24}$ while the
other $\Zar{0}{0}$ terms can be combined with the $\Zar{0}{\hd}$,
$\Zar{\hd}{0}$, and $\Zar{\hd}{\hd}$ shifted lattice sums
to yield projected sums where $\d$ runs from $0$ to $2^{d}-1$:
\begin{equation}
\label{xp}
\Xi_{(F_{\d_1})^4}
= 96 \eta^{24} \Zar{0}{0}+ \left[
 \left( \frac{\th_3^{''''}}{\th_3} +\frac{\th_4^{''''}}{\th_4}
\right) \th_3^8 \th_4^8 \Zar{0}{\d} \orb \right] \ .
\end{equation}
We can now use the results of Appendix \ref{lat} to compute the
integral of \eqref{xp} on the fundamental domain ${\cal F}$ of
$Sl(2,\Zint)$. For this, we note that the holomorphic form
$\eta^{24}$ in the first term cancels against the BPS partition
function $1/\eta^{24}$ in \eqref{delta}. As far as the terms in
brackets is concerned, the integral on ${\cal F}$ can be unfolded
on the fundamental domain ${\cal F}_2^-$ of the $\Gamma_2^-$
congruence subgroup  of $Sl(2,\Zint)$, a three-fold cover of
${\cal F}$, by keeping the first term only. According to the
property stated in \eqref{thm}, and using \eqref{h5} we can then
replace the modular form
$(\th_3^{''''}/\th_3+\th_4^{''''}/\th_4)\th_3^8\th_4^8/\eta^{24}$
by two thirds its value under the Hecke operator \eqref{hg2-}
which turns out to be zero in this case. We finally obtain
\begin{equation}
\label{d4p} \Delta_{(F_{\d_1})^4} =  \frac{96}{4!\cdot 2^{d+1}}
\int_{\F} \frac{d^2\tau}{\tau_2^2} \Zar{0}{0} \ ,
\end{equation}
which we recall is the complete expression in the Abelian case
$d=4$. The fact that all oscillator contributions have cancelled
will be crucial for heterotic type II duality to hold, as we will
discuss in Section \ref{dual}.

Moving on to the expression in \eqref{xi22}, the manipulations are
identical and making use of the summation identity \eqref{s5} and
of the ``Hecke'' identity \eqref{h6}, yield
\begin{equation}
\label{d22} \Delta_{(\Tr F_{\d_1}^2)^2} = \frac{32}{3\cdot 8\cdot
2^{d+1}} \int_{\F_2^-} \frac{d^2\tau}{\tau_2^2}  \Zar{0}{\hd} \
,\quad (d\leq 3) \ .
\end{equation}
Combining \eqref{d4p} and \eqref{d22} together, we thus obtain the
full $\Tr F_{\d}^4$ coupling for a non-Abelian gauge group,
\begin{equation}
\label{d4} \Delta_{\Tr F_{\d_1}^4}
 = \frac{32}{4!\cdot 2^{d+1}} \int_{\F_2^-}
\frac{d^2\tau}{\tau_2^2}  \left( 2 \Zar{0}{0} - \Zar{0}{\d}
\right) \ ,\quad (d \leq 3) \ .
\end{equation}

\subsubsection{$\Tr F_{\d_1}^2 \Tr F_{\d_2}^2 $ \label{hetaf22}}
Considering now the coupling between two different gauge groups,
we get for $d>1$
\begin{align}
\label{xi22p} \Xi_{\Tr F_{\d_1}^2 \Tr F_{\d_2}^2 }
= & \th_\alpha^{14}(\th_\alpha^{''})^2 Z\ar{0}{0}
\nn \\
 & + \left[ \left( 2
\frac{\th_3^{''}}{\th_3} \frac{\th_4^{''}}{\th_4} Z\ar{00}{\d1} +
\left( \left(\frac{\th_3^{''}}{\th_3}\right)^2+
 \left(\frac{\th_4^{''}}{\th_4}\right)^2\right)  Z\ar{00}{\hd 0}
\right) \th_3^8\th_4^8 \orb \right] \ .
\end{align}
Here, $\d$ in the second term runs over the $(d-1)$-digit binary
numbers (zero included), whereas $\hd $ runs over the $(d-1)$-digit
binary numbers in the last term (zero excluded). Here we have made
a particular choice of gauge fields $F_{\d_1},F_{\d_2}$
corresponding to Wilson lines $y_{\d_1}={\rm bin}(0),y_{\d_2}={\rm
bin}(1)$, but the other amplitudes can be obtained by T-duality,
and the structure in \eqref{xi22p} is generic. For $d=1$, the
second term does not make sense, and we have instead
\begin{equation}
\label{xi22p1}
\Xi_{\Tr F_{\d_1}^2 \Tr F_{\d_2}^2 }
=\th_\alpha^{14}(\th_\alpha^{''})^2 Z_{1,1}\ar{0}{0}+ \left[ 2
\frac{\th_3^{''}}{\th_3} \frac{\th_4^{''}}{\th_4}
 Z_{1,1}\ar{0}{1} \th_3^8\th_4^8 \orb \right]\quad (d=1) \ .
\end{equation}
The simplification of expression \eqref{xi22p} is more involved
than the ones of the previous subsection. In this case, we use
the summation identity \eqref{s4} along with \eqref{t10} to bring the
first and last term in the form of the middle one,
\begin{equation}
\label{xi22pb} \Xi_{\Tr F_{\d_1}^2 \Tr F_{\d_2}^2 } = 16 \eta^{24}
Z\ar{0}{0}+ \left[ \left( 2 \frac{\th_3^{''}}{\th_3}
\frac{\th_4^{''}}{\th_4} Z\ar{0}{\d} + \frac{1}{16} \th_2^8
Z\ar{00}{\hd 0} \right) \th_3^8\th_4^8 \orb \right] \ .
\end{equation}
Then, using the identity $\th_2\th_3\th_4=2\eta^3$, we see
that the first term and the second in the bracket
are proportional to $\eta^{24}$, while we can
use the Hecke identity \eqref{h7} for the middle term arriving at
\begin{subequations}
\label{d22p}
\begin{equation}
\Delta_{\Tr F_{\d_1}^2 \Tr F_{\d_2}^2 }
= \frac{16}{3\cdot 4 \cdot 2^{d+1}} \int_{\F_2^-}
\frac{d^2\tau}{\tau_2^2}  \left(
\Zar{0}{0} - \Zar{0}{\d} +  3 \Zar{00}{\hd 0} \right)
\end{equation}
\begin{equation}
 = \frac{16}{3\cdot 4\cdot 2^{d+1}} \int_{\F_2^-}
\frac{d^2\tau}{\tau_2^2} \left( 2Z\ar{0}{\hd} - 3 Z\ar{00}{\d 1}
\right) \quad (d > 1)  \ .
\end{equation}
\end{subequations}
Similarly, for the special case $d=1$ in \eqref{xi22p1}
we find after analogous steps
\begin{equation}
\label{d22p1}
 \Delta_{\Tr F_{\d_1}^2 \Tr F_{\d_2}^2 } = -\frac{16}{3\cdot 4\cdot 4}
\int_{\F_2^-} \frac{d^2\tau}{\tau_2^2} Z_{1,1}\ar{0}{1}\quad (d=1) \ .
\end{equation}

\subsubsection{$F_{\d_1}F_{\d_2}F_{\d_3}F_{\d_4}$ \label{hetaf1111}}
For $d<4$, the gauge lattice partition function \eqref{shiftt}
is even under $v_I^\d\to-v_I^\d$, so that such a term cannot occur,
in agreement with the fact that the generators of $SO(2^{5-d})$ are
traceless. For $d=4$ however, it turns out that the coupling
between four different $U(1)$ does not vanish, provided the
selection rule
\begin{equation}
\label{selrule}
\d_1+\d_2+\d_3+\d_4 = 0 \mod 16
\end{equation}
is obeyed, in which case
\begin{align}
\label{xi1111}
 \Xi_{F_{\d_1}F_{\d_2}F_{\d_3}F_{\d_4}} =&
\eta^{12}(\th_2\th_3\th_4)^4 \cdot \\ &\cdot \left(
\Zar{0100}{1000}+\Zar{1000}{0100}+\Zar{0100}{1100}+
\Zar{1000}{1100}+\Zar{1100}{0100}+\Zar{1100}{1000} \right) \ .
\nonumber
\end{align}
Note that the modular orbit now involves six different shifted
partition functions. The precise orbit depends on the choice of the
four $U(1)$, and we have chosen one example corresponding to
$y_{\d_1}={(0000)},y_{\d_2}={(0001)},y_{\d_3}={(0010)},y_{\d_4}={(0011)}$.
Using the relation $\th_2\th_3\th_4=2\eta^3$, we see that the
modular form again cancels against the partition function of the
half-BPS
states\footnote{The same mechanism was observed in \cite{Gava:1999ky}.},
and we are therefore left with
\begin{equation}
\label{f41111}
 \Delta_{F_{\d_1}F_{\d_2}F_{\d_3}F_{\d_4}} =  \frac{16}{2^{5}}
\int_{{\cal F}_2}
\frac{d^2\tau}{\tau_2^2} \Zar{0100}{1000}\quad (d=4) \ ,
\end{equation}
where the integration is over the six-fold cover $\F_2$ of
the fundamental domain $\F$ of $Sl(2,\Zint)$.
As we will see in Section \ref{dualii}, the selection rule \eqref{selrule}
has a direct counterpart in the dual  type IIA theory.

\subsubsection{$\Tr R^4$, $(\Tr R^2)^2$ and $\Tr R^2 \Tr F_{\d}^2$
\label{hetar4}}
We now turn to four-point functions involving gravitons.
For a four-graviton amplitude, the elliptic genus \cite{Lerche:1988zy} is as
in the uncompactified heterotic theory, and yields
\begin{subequations}
\begin{equation}
\Xi_{\Tr R^4}=\frac{E_4}{2^7\cdot 3^2\cdot 5} \left[
\theta_{\alpha}^{16} Z\ar{0}{0}+ 2\left( \th_3^{8}\th_4^{8}
Z\ar{0}{\hd} \orb \right)  \right] \ ,
\end{equation}
\begin{equation}
\label{xigg} \Xi_{(\Tr R^2)^2}=\frac{\et^2}{2^9\cdot 3^2} \left[
\theta_{\alpha}^{16} Z\ar{0}{0}+ 2\left( \th_3^{8}\th_4^{8}
Z\ar{0}{\hd} \orb \right)  \right] \ .
\end{equation}
\end{subequations}
For two-graviton two-gauge-boson scattering on the other hand,
we need to take two derivatives with respect to $v_\d$, and we get
\begin{equation}
\label{xir2f}
 \Xi_{\Tr R^2\Tr F_{\d_1}^2}=\frac{\et}{2^3\cdot 3}
\left[ \th_\alpha^{15} \th_\alpha'' Z\ar{0}{0} +
\left(\frac{\th_3''}{\th_3}+\frac{\th_4''}{\th_4}\right)
\th_3^{8}\th_4^{8} Z\ar{0}{\hd} \orb   \right] \ .
\end{equation}

These amplitudes can all be simplified again by the now familiar method.
In particular using \eqref{s1}, \eqref{h2} we obtain
\begin{subequations}
\label{r4}
\begin{equation}
\label{r41}
\Delta_{\Tr R^4}= \frac{480}{2^7\cdot 3^2\cdot 5 \cdot 2^{d+1}}
 \int_{\F_2^-}
\frac{d^2\tau}{\tau_2^2}   \Zar{0}{\d} \ ,
\end{equation}
\begin{equation}
\label{r22}
\Delta_{(\Tr R^2)^2}= \frac{96}{2^9\cdot 3^2\cdot 2^{d+1}}
  \int_{\F_2^-}
\frac{d^2\tau}{\tau_2^2}  \Zar{0}{\d} \ .
\end{equation}
\end{subequations}
It is worthwhile noting that $\Delta_{\Tr R^4}=4 \Delta_{(\Tr R^2)^2}$:
this implies that the two terms can be combined into a $t_8 t_8 R^4=
t_8 (4 \Tr R^4 - (\Tr R^2)^2)$ coupling, as also arises in type IIA
on $K_3$ \cite{Berkovits:1998ex}.
Using \eqref{s2}, \eqref{h4} in \eqref{xir2f} we similarly find
for the coupling of two gravitons and two right moving $(0,16)$
gauge fields,
\begin{equation}
\label{drf} \Delta_{\Tr R^2\Tr F_{\d_1}^2}=\frac{16}{2^3\cdot
3\cdot 2^{d+1}}
  \int_{\F_2^-} \frac{d^2\tau}{\tau_2^2}   \Zar{0}{\d}\ .
\end{equation}

\subsubsection{$(0,d)$ gauge bosons \label{heta0d}}
As we mentioned above, the insertion of momenta are more easily
dealt with in the Lagrangian representation. We thus get
\begin{equation}
\Xi_{F_i F_j F_k F_l }= q_R^i q_R^j q_R^k q_R^l
\left[
\theta_{\alpha}^{16} Z\ar{0}{0}+ 2\left(
\th_3^{8}\th_4^{8} Z\ar{0}{\hd} \orb \right)  \right] \ ,
\end{equation}
where $q_R^i$ acts on the torus partition function by inserting a
factor of $(m^i+d^i/2-\tau (n^i+d^{'i}/2))/\tau_2$ for
$Z\ar{\d'}{\d}$. We can also consider the mixed amplitudes of two
$(0,d)$ gauge bosons and two gauge bosons or two gravitons
respectively, for which \begin{subequations}
\begin{equation} \Xi_{(F_i F_j) \Tr R^2 }=
q_R^i q_R^j \frac{\et}{2^3\cdot 3}
 \left[
\theta_{\alpha}^{16} Z\ar{0}{0}+ 2\left( \th_3^{8}\th_4^{8}
Z\ar{0}{\hd} \orb \right)  \right] \ ,
\end{equation}
\begin{equation}
\Xi_{(F_i F_j) \Tr F_{\d_1}^2 } =q_R^i q_R^j \left[ \th_\alpha^{15}
\th_\alpha'' Z\ar{0}{0} +
\left(\frac{\th_3''}{\th_3}+\frac{\th_4''}{\th_4}\right)
\th_3^{8}\th_4^{8} Z\ar{0}{\hd} \orb   \right] \ .
\end{equation}
\end{subequations}
Using the methods described above and \eqref{s1}, \eqref{h1} or
\eqref{s2}, \eqref{h3}, the corresponding couplings are all zero,
and we record
\begin{equation}
\label{f0} \Delta_{F_i F_j F_k F_l } =\Delta_{(F_i F_j) \Tr R^2
}=\Delta_{(F_i F_j) \Tr F_\d^2 } = 0 \ .
\end{equation}

\subsubsection{Graviphotons and summary \label{hetagp}}
We can simply obtain the amplitudes with graviphotons by noting
that the graviphoton and graviton vertex operators are similar.
We thus obtain four powers of momenta on the holomorphic side,
and four on the antiholomorphic side, so that the four
graviphoton amplitude starts at eight-derivative order only.
In the case of two graviphotons and two right-moving gauge bosons,
we obtain two extra powers of momenta from the antiholomorphic
side. As a result, the amplitude starts at six-derivative order only.
All $F^4$ effective couplings involving graviphotons thus vanish.

We can summarize the above as follows.
At a generic point of the moduli space, and at heterotic one-loop:

\noindent (i)
The first non-trivial correction to the $(0,d+16)^4$
couplings occurs at the four-derivative level. \newline
\noindent (ii)
The first non-trivial correction to the $(d,0)^4$ couplings
occurs at the eight-derivative level. \newline
\noindent (iii)
The first non-trivial correction to the $(d,0)^2(0,d+16)^2$
couplings occurs at the six-derivative level.

At the $\Zint_2$-orbifold point of the six-dimensional heterotic
string:
\newline
\noindent (a) The four-derivative $(0,16)^4$ couplings have
non-vanishing one-loop corrections. \newline \noindent (b)  The
one-loop four-derivative $(4,4)^4$ and $(4,4)^2(0,16)^2$ couplings
are vanishing. The first non-trivial correction for these
couplings occurs at the eight-derivative level.

\section{Heterotic-type IIA duality in six dimensions \label{dual}}
As we already argued in the introduction, the $F^4$ couplings 
in the heterotic string on $T^4$ are
given at one-loop only, and translate,
through the standard duality map, into a purely
tree-level coupling in type IIA on $K_3$. We can therefore perform
a very quantitative test of heterotic-type IIA duality by
computing the tree-level $F^4$ amplitude on the type II side at an
orbifold point, where the CFT is exactly soluble. This is the
object of the first subsection, the results of which will be
summarized and compared to the heterotic side in the second. The
reader appalled by the technicalities of Section \ref{dualii}
should not feel guilty in proceeding to Section \ref{dualt}.

\subsection{Type IIA four gauge field amplitude \label{dualii}}
The 24 gauge fields in type IIA on $K_3$ originate from the
Ramond-Ramond sector. The 4 graviphotons can be understood as the
reduction $G_i$ of the  4-form field strength in $D=10$ on the 3
self-dual cycles of $K_3$ together with the ten-dimensional 2-form
field-strength $G_0$, whereas the 20 vector multiplets come from
the 4-form in $D=10$ on the 19 anti-self-dual cycles together with
the 6-form field strength $K_3$ itself: we denote them by $F_I$
and $F_0$ respectively.

\subsubsection{Vertex operators}
At the $T^4/\Zint_2$ orbifold point, the gauge fields split into
untwisted and twisted sectors. The untwisted sector contributes
$(4,4)$ of them, whose vertex operators are simple projections of
the ten-dimensional Ramond vertex
\cite{Friedan:1985ey,Cohn:1986bn}, and can be decomposed into a
product of $SO(6)$ and $SO(4)$ spin fields times a ghost part. The
$SO(6)$ spin fields $S_\alpha(z), \bar S_\alpha(\bar z)$ have
conformal dimension $3/8$ and transform as an $SO(6)$ spinor of
positive chirality, while $S^\alpha$ and $\bar S^{\alpha}(\bar z)$ are
$SO(6)$ spinors of negative chirality. The $SO(4)$ spin-fields $\Psi_a$ and
$\Psi_{\ad}$
involve both chiralities, and are most easily described using the standard
dotted notation for $SO(4)\sim SU(2)\times SU(2)$. The ten-dimensional $SO(10)$
spinor decomposes under $SO(4)\times SO(6)$ as $({\bf 2,4})+({\bf \bar 2,\bar 4})$,
while the $SO(10)$ conjugate spinor decomposes as $({\bf 2,\bar 4})+({\bf 2,\bar 4})$.
The orbifold projection on $T^4$ acts on the $SO(4)$ spinors as $({\bf 2,\bar 2})\to ({\bf 2},-{\bf\bar 2})$.
Hence, the vertex
operators for untwisted gauge fields read
\begin{subequations}
\begin{eqnarray}
\label{ut} V_{-1/2}^{m}&=&e ^{-\phi/2}e ^{-\tilde \phi/2}~ X^m
~S_{\alpha} \tilde S^{\beta}~ {\Sigma^{\mu \nu} }^{\alpha}_{\beta}
~\zeta_{\mu \nu} ~e^{ikX} \ , \\
\bar V_{-1/2}^{m}&=&e ^{-\phi/2}e
^{-\tilde \phi/2}~ \bar X^m ~S^{\alpha} \tilde S_{\beta}
~{\Sigma^{\mu \nu}}_{\alpha}^{\beta}~\zeta_{\mu \nu}~ e^{ikX} \ ,
\end{eqnarray}
\end{subequations}
where we use the covariant formalism of \cite{Friedan:1985ey}.
$\Sigma^{\mu \nu}_{\alpha\beta}$ are the $SO(6)$ rotation matrices
in the spinor representation, and $\zeta_{\mu\nu}$ the polarisation
tensors of the field strengths.
$e^{-\phi/2}$ is the bosonized superconformal ghost of conformal
dimension $3/8$. $X^m, \bar X^m$ are the fermion combinations
\begin{equation}
X^{m}=( \Psi^{\a}\epsilon_{\a\b}\tilde \Psi^{\b}\ ,
\Psi^{\a}{\sigma^{ij}}_{\a\b}\tilde \Psi^{\b})\ ,
\quad \bar
X^{m}=( \Psi^{\ad}\epsilon_{\ad\bd}\tilde \Psi^{\bd}\ ,
\Psi^{\ad}{\bar\sigma^{ij}}_{\ad\bd}\tilde \Psi^{\bd})\ ,
\end{equation}
where $m$ runs from 0 to 3 and $\sigma^m=(i1_2,\sigma^i)$.
Because of self-duality, only 3 components
of $\sigma^{ij}$ contribute. Together,
$X^m$ and $\bar X^m$ transform as a conjugate spinor of the
T-duality group $SO(4,4)$. We shall refer to the gauge fields
with vertex operators $V$ and $\bar V$ as chiral and antichiral
respectively. The vertex operators have been displayed
in the $(-1/2,-1/2)$ ghost picture, as appropriate for a
tree-level four-point amplitude. The 16 remaining gauge bosons
come from the 16 twisted sectors, and their vertex operators
involve twist fields $H^I$, $I=1\dots 16$ of conformal dimension
1/4,
\begin{equation}
\label{tw} V_{-1/2}^{T}= e ^{-\phi/2}e ^{-\tilde \phi/2}~ H^I
~S^{\alpha} \tilde S_{\beta}~ {\Sigma^{\mu \nu}}_{\alpha}^{\beta}~
\zeta_{\mu \nu} \ .
\end{equation}
We have omitted the momentum part, since the Ramond-Ramond vertex
operators couple to the world-sheet only through their
field-strength, which already provides the 4 necessary
derivatives. We now consider a tree-level amplitude with four
vertex operators inserted at $0,x,1,\infty$ on the complex plane.
The correlator factorizes into the ghost part,
\begin{equation}
\label{ghost} \langle e ^{-\phi/2}(\infty)e ^{-\phi/2}(1)\ e
^{-\phi/2}(x) e ^{-\phi/2}(0) \rangle = [x(1-x)]^{-1/4}\ ,
\end{equation}
a 6D spin field part,
\begin{subequations}\label{spin}
\begin{equation}
 \langle S_\alpha(\infty) S_\beta(1)\ S_\gamma(x)
S_\delta(0) \rangle = [x(1-x)]^{-1/4}
\epsilon_{\alpha\beta\gamma\delta} \ ,
\end{equation}
\begin{equation}
 \langle S_\alpha(\infty) S^\beta(1)\ S_\gamma(x)
S^\delta(0) \rangle = \left[ \delta_\a^\b \delta_\gamma ^\delta x +
\delta_\a^\delta \delta_\gamma ^\b (1-x) \right]
[x(1-x)]^{-3/4}
\end{equation}
\end{subequations}
and an internal part which depends on the gauge bosons of
interest. Equations \eqref{spin} are easily obtained by bosonizing
the spin-fields along the lines of \cite{Cohn:1986bn}, and show
that we already get the correct kinematical structure,
\begin{equation}
\epsilon_{\alpha\beta\gamma\delta}
\epsilon_{\bar\alpha\bar\beta\bar\gamma\bar\delta}
(\Sigma_{\alpha\bar\alpha}\cdot \zeta)
(\Sigma_{\beta\bar\beta}\cdot \zeta)
(\Sigma_{\gamma\bar\gamma}\cdot \zeta)
(\Sigma_{\delta\bar\delta}\cdot \zeta) = t_8 \zeta^4\ .
\end{equation}

\subsubsection{Four-twist-field amplitude}
We now consider more specifically the amplitude between four gauge
bosons from the twisted sector. The internal part is given by the
correlator of four twist fields on $T^4/\Zint_2$. Since twist
fields create a $\Zint_2$ cut in the world-sheet, this is
equivalent to a vacuum amplitude on the covering surface of the
sphere with 4 punctures, namely a genus one surface
\cite{Dixon:1985jw}. This equivalence will turn out to be crucial
for the heterotic-type II duality to hold. More precisely, the
modular parameter of the torus is related to the vertex positions
through the Picard map
\begin{equation}
x=\left(\frac{\th_3}{\th_4}\right)^4\ ,\quad \frac{dx}{x}=i\pi
\th_2^4 d\tau\ ,
\end{equation}
so that the four-point amplitude for $T^4/\Zint_2$ twist fields is
given by a slight adaptation from \cite{Dixon:1987qv},
\begin{equation}
\label{twist} \langle H_{\epsilon_1}(\infty) H_{\epsilon_2}(1)
H_{\epsilon_3}(x) H_{\epsilon_4}(0) \rangle = 2^{-8/3}
\frac{Z_{4,4}
\ar{\epsilon^1_i+\epsilon^4_i}{\epsilon^1_i+\epsilon^2_i}}
{\tau_2^2 \eta^4 \bar \eta ^4} |x(1-x)|^{-1/3}
\end{equation}
if charge conservation
$\epsilon^1_i+\epsilon^2_i+\epsilon^3_i+\epsilon^4_i =0 \mod 2$ is
obeyed for every $i=1 \dots 4$, zero otherwise. This selection
rule results from a discrete group of symmetries of the orbifold
CFT \cite{Dijkgraaf:1988vp}, which correspond to half lattice
translations on the covering torus $T^4$, as well as their T-dual
counterparts. The four translations exchange the 16 twisted
sectors in pairs, while the T-dual translations act by $-1$ on
eight of the 16 twisted sectors. These symmetries commute up to a
global $-1$ factor on all twisted sectors, and thus generate a
dihedral group $\Zint_2 \ltimes \Zint_2^8 $ which generalizes
the $D_4$ symmetry of the $S_1/\Zint_2$ orbifold CFT. The above
selection rule is precisely the one encountered on the heterotic
side \eqref{selrule}, providing new support for the duality.

Putting
\eqref{twist} together with \eqref{ghost} and \eqref{spin}, and changing
the integration variable from $x$ to $\tau$, we therefore obtain
\begin{equation}
\label{a4iig} \Delta=\frac{\lii^6}{V_{K_3}}
\int_{{\cal F}_2}  \frac{d^2\tau}{\tau_2^2}
Z_{4,4}\ar{\epsilon^1_i+\epsilon^4_i}{\epsilon^1_i+\epsilon^2_i} 
(G/\lii^2,B) \ ,
\end{equation}
where we dropped an overall constant. The integration runs over
the fundamental domain of the index 6 subgroup of $Sl(2,\Zint)$,
which is the moduli space of the sphere with 4 punctures
(see appendix A.2 for a discussion of congruence 2 subgroups
of $Sl(2,\Zint)$). Note in
particular, that the oscillators in \eqref{twist} have dropped, in
agreement with the fact that this amplitude should be half-BPS
saturated. The normalization factor $\lii^6/V_{K_3}$  has been
chosen so as to agree with the heterotic result.

It is useful to discuss more specifically which shifts occur in
the lattice partition function. Firstly, we note that a
permutation of the four twist fields can be re-absorbed by a
modular transformation which maps the extended fundamental domain
${\cal F}_2$ to itself, and hence leaves the integral invariant.
We thus have only three possible results, up to permutations of
the torus directions: 

\noindent (i) if all twist fields sit at the same point,
\begin{equation}
\label{a4ii} \Delta_{(F_I)^4} =\frac{\lii^6}{V_{K_3}}
\int_{{\cal F}_2} \frac{d^2\tau}{\tau_2^2}
Z_{4,4}\ar{0000}{0000} = 6 \frac{\lii^6}{V_{K_3}}
\int_{\cal F} \frac{d^2\tau}{\tau_2^2}
Z_{4,4}\ar{0000}{0000}\ ,
\end{equation}
(ii) if they are separated in two pairs,
\begin{equation}
\label{a22ii} \Delta_{(F_I)^2(F_J)^2}=\frac{\lii^6}{V_{K_3}}
\int_{{\cal F}_2} \frac{d^2\tau}{\tau_2^2}
Z_{4,4}\ar{0000}{0001} = 3\frac{\lii^6}{V_{K_3}}\int_{{\cal F}_2^-}
\frac{d^2\tau}{\tau_2^2} Z_{4,4}\ar{0000}{0001}\ ,
\end{equation}
(iii) if they sit at different fixed points, yet satisfying the
selection rule,
\begin{equation}
\label{a1111ii} \Delta_{F_I F_J F_K F_L}=\frac{\lii^6}{V_{K_3}}
\int_{{\cal F}_2}
\frac{d^2\tau}{\tau_2^2} Z_{4,4}\ar{0100}{1000} \ .
\end{equation}
Again, the precise shifts appearing in
\eqref{a22ii},\eqref{a1111ii} depend on the choice of twist
fields, but the orbit structure is general. In the above
amplitudes, we have implicitly subtracted the infrared divergence
coming from the vacuum in the lattice partition functions, which
from the point of view of the tree-level amplitude correspond to
the exchange of massless particles.

\subsubsection{Four-untwisted-field amplitude}

In contrast to the previous case, the correlator between four
untwisted fields does not involve the covering torus, and we have
to deal with a genuine tree-level computation. The computation of
various scattering amplitudes of four gauge bosons is then identical 
to the analogous computation in the maximally
supersymmetric type II theory. This computation has not been done
to our knowledge but a quick argument already indicates that the
four derivative couplings of (4,4) gauge bosons vanish at tree
level. Indeed, the leading corrections to gravitational couplings
occur at the 8-derivative level \cite{Grisaru:1986dk}. By
supersymmetry, we expect that non-trivial corrections to
Ramond-Ramond self-couplings should start at the eight-derivative
level as well. Since Ramond-Ramond fields in ten dimensions
descend to the (4,4) gauge fields upon compactification to 6
dimensions, it is evident that there should be no
(four-derivative) $F^4$ terms for these fields. This of course
does not preclude the existence of $F^4$ couplings mixing twisted
and untwisted gauge fields.

The correlators of
$SO(4)$ spin fields can be simply obtained by the usual
bosonization techniques, and read
\begin{subequations}
\begin{eqnarray}
\langle \Psi_{\a}(\infty)\Psi_{\b}(1)\Psi_{\c{}}(x)\Psi_{\delta}(0)
\rangle &=& \frac{1}{ \sqrt{x(1-x)}}
\left[\e_{\a\b}\e_{{\c{}}\delta}-
x~\e_{\a{\c{}}}\e_{\b\delta}\right]\ , \\
\langle\Psi_{\a}(\infty)\Psi_{\bd}(1)\Psi_{\c{}}(x)\Psi_{\dd}(0)\rangle
&=&\e_{\a{\c{}}}\e_{\bd\dd}\ ,
\end{eqnarray}
\end{subequations}
and similarly for the right-moving fields.
Including the contribution from the ghost, 6D spin fields, as well
as the momentum dependence $|x|^{-s/4}~ |1-x|^{-t/4}$ with
$k_2\cdot k_3=-t/2$, $k_3\cdot k_4=-s/2$, $s+t+u=0$, the total
amplitude reduces to a combination of standard integrals
\begin{equation}
\int d^2 x |x|^{-a-s/4}|1-x|^{-b-t/4}=\pi
\frac{\Gamma\left(1-\frac{a}{2}-\frac{s}{8}\right)
      \Gamma\left(1-\frac{b}{2}-\frac{t}{8}\right)
     \Gamma\left(\frac{a+b}{2}-1-\frac{u}{8}\right)}
{ \Gamma\left(2-\frac{a+b}{2}+\frac{u}{8}\right)
  \Gamma\left(\frac{a}{2}+\frac{s}{8}\right)
  \Gamma\left(\frac{b}{2}+\frac{t}{8}\right)}
\end{equation}
with $(a,b)=(0,2),(2,0)$ and $(2,2)$ in the $s,t,u$-channels
respectively. Expanding a typical contribution for small
momenta, we have
\begin{equation}
\label{aaaa} A(s,t)= \frac{\Gamma(1-s/8)\Gamma(-t/8)\Gamma(-u/8)}{
\Gamma(1+u/8)\Gamma(s/8)\Gamma(1+t/8)} =\frac{s^2}{8} \left( \frac{1}{stu}+
\frac{\zeta(3)}{256} + \dots \right) \ .
\end{equation}
The pole term corresponds to the tree-level massless exchange, and
has to be subtracted in order to extract a correction to the
effective action as in the twisted case. The correction only
occurs at order $s^2$, corresponding to an eight-derivative
coupling in the effective action. Hence there are no $F^4$
couplings at tree-level between four untwisted chiral fields, nor
between four antichiral fields, as we anticipated at the beginning
of this section. The first non-trivial correction however implies
$\pa^4 F^4$ couplings, which are nothing but the ten-dimensional
eight-derivative couplings, related by supersymmetry to the $t_8
t_8 R^4$ couplings \cite{Grisaru:1986dk}, reduced on the torus
$T^4$.

There is a puzzle concerning the $(4,0)^2(0,16)^2$ threshold. In
the heterotic string this was shown to vanish. A type II
computation along the lines above seems to give a non-zero answer.
Clearly this deserves further study.

\subsection{Duality and triality \label{dualt}}
Let us summarize the salient features of our computations so far,
concentrating on the simple case of four-(0,16) gauge boson
scattering for now.
\begin{itemize}
\item On the heterotic side, the one-loop amplitude
was expressed as the integral over the fundamental domain of the
lattice partition function of the torus $T^4$ with particular
shifts, with an insertion of an elliptic genus
$\Phi(\tau)=(\alpha E_4 + \beta \et^2) \th_3^8\th_4^8/\eta^{24}$.
This structure is characteristic of half-BPS heterotic couplings, where
the fermionic zero-modes are just saturated on the left-hand side
and the right-moving oscillators generate the Hagedorn density of
BPS states. Thanks to the Hecke identities described in Appendix
\ref{moduh}, the elliptic genus has dropped, leaving a simple
integral of a shifted lattice partition function such as
\eqref{d4p},\eqref{d22p} and \eqref{f41111}.  
A particular selection rule \eqref{selrule} was
also found.

\item On the type IIA side, the tree-level amplitude of four twist fields
has turned out to secretly be a genus 1 amplitude \eqref{a4iig} on
the covering of the 4-punctured sphere. The BPS nature of the
coupling was revealed in the cancellation of the bosonic and
fermionic determinants on the covering surface. The selection rule
was a simple consequence of the $\Zint_2 \ltimes  \Zint_2^8$
discrete symmetries of the orbifold. Eventually, the amplitude
reduced to an integral \eqref{a4iig} of the shifted partition function of the
torus covering the $K_3$ surface, in agreement with the heterotic
results \eqref{d4p}, \eqref{d22p}, \eqref{f41111}. 
The tree-level amplitude for untwisted
fields on the other hand was shown in \eqref{aaaa} to vanish at 4
derivative order for $(0,d)$ or $(d,0)$ gauge bosons, in
accordance with the heterotic result \eqref{f0}\footnote{As
mentioned before the issue of matching the $(0,4)^2(0,16)^2$
threshold remains obscure.}.
\end{itemize}

However, it takes yet another miracle to identify the heterotic
result with the type II result: indeed
the two tori of moduli $(g,b)$ and $(G,B)$ are not identical, but,
as we argued in Section \ref{modii}, related by triality,
\begin{equation}
\label{trial3} V_{K_3}=R_1^2\ ,\quad \gamma=g_3\ ,\quad
\beta=B_{33}\ ,\quad B_3 = A\ , \quad B_{\bar 3}=B_{13} \ .
\end{equation}
This transformation is certainly not a symmetry of the integrand
$Z_{4,4}\ar{0}{0}$, as a simple study of various
decompactification limits makes clear. However, it has been
shown in \cite{Obers:1999um,Obers:1999es} that the integrated result could be
rewritten as an Eisenstein series in either the vector, spinor or
the conjugate spinor representation,
\begin{equation}
\label{trial4} \int_{\cal F}
\frac{d^2\tau}{\tau_2^2}Z_{4,4}\ar{0}{0} =\frac1\pi
\eis{SO(4,4,\Zint)}{V}{s=1} =\frac1\pi\eis{SO(4,4,\Zint)}{S}{s=1}
=\frac1\pi \eis{SO(4,4,\Zint)}{C}{s=1} \ ,
\end{equation}
which implies the identity of the heterotic and type II
results,
\begin{equation}
\label{trial5} \int_{\cal F}
\frac{d^2\tau}{\tau_2^2}Z_{4,4}\ar{0}{0}(g,b) =\int_{\cal F}
\frac{d^2\tau}{\tau_2^2}Z_{4,4}\ar{0}{0}(G,B) \ .
\end{equation}
This claim was supported in \cite{Obers:1999um} by showing that
either of the terms in \eqref{trial4} was an eigenmode of the
Laplacian operator on \so{4}{4}, and of another non-invariant
second order differential operator as well; it was also shown that
the large volume and decompactification limits agreed. The same
arguments can also be made for the two terms of \eqref{trial5}
without using Eisenstein series as an intermediate step. While
heterotic-type II duality clearly implies \eqref{trial5}, it
would be useful to have a mathematically rigorous proof for it.

In
the case of four identical gauge fields, the identity \eqref{trial4}
directly matches the heterotic
result \eqref{d4p} with the corresponding type II result \eqref{a4ii}. More
generally, we need an extension of this identity
to the case with half-integer shifts. This can be obtained by
re-expressing the shifted lattice sums as unshifted lattice sums
with redefined moduli, and apply the triality \eqref{trial4}.
For example, we may rewrite the heterotic amplitude
\begin{subequations}
\begin{align}
\int_{{\cal F}_2^-} Z_{4,4}\ar{0}{\d}(g,b)&=
\frac{12}{\pi} \eis{SO(4,4,\Zint)}{V}{s=1}(g/2,b/2)
= \frac{12}{\pi}\sum_{m_i\in 2\Zint,n^i\in\Zint} \frac{2}
{{\cal M}^2_{\irrep{V}}(g,b)} \\
&= \frac{12}{\pi}
  \sum_{m,m_{ij}\in 2\Zint,m^{1j},n\in\Zint} \frac{2}
{{\cal M}^2_{\irrep{C}}(G,B)}
=\frac{12}{\pi} \eis{SO(4,4,\Zint)}{C}{s=1}(R_1/2)\\
&=\int_{{\cal F}_2^-} \left( 6 Z_{4,4}\ar{0000}{1000}(G,B)+ 2
Z_{4,4}\ar{0000}{0000}(G,B) \right)
\end{align}
\end{subequations}
in a form suitable for comparison with type II amplitudes.
Here, we have used \eqref{mi4}, \eqref{I4} in the first step to
convert to an Eisenstein series in the vector representation, in the
 second step we have rewritten this series as a constrained Eisenstein
series involving the vector mass at the original heterotic moduli.
The third step consists of the application of the triality map
\eqref{trial1}, \eqref{trial2} to write the vector mass as a
conjugate spinor mass with type II moduli, which is re-expressed
as a conjugate spinor Eisenstein series in the fourth step. Then,
the fifth step uses again \eqref{I4}, \eqref{mi4} to present the
result as a tree-level type II amplitude of the form
\eqref{a4iig}. However, the precise matching will require the
exact identification of the gauge fields on the type II side with
those on the heterotic side, which we have not been able to
achieve.
It would be also interesting to understand how the duality
holds at other orbifold points of $K_3$, since naively
the correlator of $\Zint_n$ twist fields on the sphere involves
higher genus Riemann surfaces, albeit of a very symmetric type.

Finally, let us comment on $R^4$ gravitational couplings.
In that case, the one-loop heterotic result translates
into a two-loop contribution on the type IIA side. On the other hand,
it is known that there is a $t_8 t_8 R^4$ coupling arising
at tree-level and one-loop on the type IIA side, which translate
into a two- and three-loop contribution on the heterotic side.
It would be interesting to carefully determine the combination
of these gravitational couplings that obeys a non-renormalization
theorem, if any. It would also be very  interesting to compute
higher-derivative $R^4 F^{4g-4}$ couplings on the heterotic
side, and compare them with the topological amplitudes on the type II
side \cite{Berkovits:1998ex}.

\section{Dual interpretation of higher derivative couplings
\label{inst}}
Having reproduced the type IIA tree-level $F^4$ coupling in 6 dimensions
from the heterotic one-loop amplitude, we now would like to use
the duality map to obtain some non-trivial results in other
dimensions. This will provide further checks of duality, and at
the same time give new insights into non-perturbative effects on
the dual side.

\subsection{Type I' thresholds \label{insti}}
As discussed in Section \ref{modi}, the heterotic string on $S_1$
at the $SO(16)\times SO(16)$ point is dual to type I' with eight
D8-branes located at each of the two orientifold points. The
modular integrals of shifted lattices are quite simple to compute
using the summation identity \eqref{allsum1} together with the
modular integral  \eqref{I1} for an unshifted lattice. For
$(0,16)$ gauge fields, we  then obtain from \eqref{d4},
\eqref{d22} and \eqref{d22p1},
\begin{subequations}
\begin{equation} \label{d41i}
\Delta_{\Tr F_{\d}^4}
= \frac{1}{6} \left( 2 I_{1}(R_{\rm H})-I_{1}(R_{\rm H}/2) \right) =
\frac{\pi}{3} \frac{R_{\rm H}}{\lh^2} \ ,
\end{equation}
\begin{equation} \label{d42i}
\Delta_{\Tr F_{\d}^2 \Tr F_{\d}^2 }= - \Delta_{\Tr F_{\d_1}^2 \Tr
F_{\d_2}^2 } = -\frac{1}{6} \left( I_{1}(R_{\rm H})-2
I_{1}(R_{\rm H}/2) \right) = \frac{\pi}{3R_{\rm H}} \ ,
\end{equation}
\end{subequations}
where we reinstated the powers $\lh$ on dimensional ground.
Translating to type I' variables using \eqref{dualip}, the heterotic
thresholds translate into
\begin{equation}
\Delta_{\Tr F_{\d}^4}=\frac{\pi}{3 g_{\rm I'}l_{\rm I'}}
 \ ,\quad \Delta_{\Tr
F_{\d}^2 \Tr F_{\d}^2 }= - \Delta_{\Tr F_{\d_1}^2 \Tr F_{\d_2}^2 }
=\frac{\pi R_{\rm I'}}{3 l_{\rm I'}^2 } \ .
\end{equation}
Given the heterotic non-renormalization theorem, these couplings
should therefore be given by a disk and cylinder diagram
respectively, without further corrections. In particular, note
that the absence of factorized couplings $(\Tr F^2)^2$ at tree-level
is consistent with the fact that these couplings need (at least)
two boundaries. Moreover,
 the absence of non-perturbative corrections is consistent with
the fact that there are no half-BPS instantons in type I' in 9
dimensions. The $F^4$ couplings have been studied in
\cite{Bachas:1997mc} in the context of the duality between the
$SO(32)$ heterotic string and type I, where it was noticed that
the duality requires contributions of higher genus surfaces
$(\chi=-1,-2)$ on the type I side, due to non-holomorphic
contributions to the elliptic genus. The $SO(16)\times SO(16)$
point therefore appears to be a simpler setting to further
understand heterotic-type I duality, and this is indeed the point
where this duality can be derived from the eleven-dimensional
strong coupling dynamics of the heterotic string
\cite{Polchinski:1996df}.

We may also consider how the
gravitational couplings \eqref{r4} translate under the
duality. In that case,
\begin{equation} \label{r4i}
\Delta_{\Tr R^4}=4\Delta_{(\Tr R^2)^2}= \frac{\pi}{48} \left(
\frac{R_{\rm H}}{\lh^2} + \frac{2}{R_{\rm H}} \right)
=\frac{\pi}{48} \left( \frac{1}{g_{\rm I'}l_{\rm I'}} + 2 \frac{R_{\rm
I'}}{ l_{\rm I'}^2 } \right) \ ,
\end{equation}
so that they receive contributions both at tree-level and one-loop
on the type I' side. As already discussed, it is unclear if
the non-renormalization theorem applies on the heterotic
side, and they may therefore get contributions from higher loops.

\subsection{F-theory on $K_3$ and O7-plane interactions \label{instf}}

As discussed in Section \ref{modf}, the heterotic string on $T^2$
at the $SO(8)^4$ point is dual to type IIB on a $T^2/\Zint_2$
orientifold, which is nothing but F-theory on $K_3$ at the
orbifold point. The $F^4$ and related couplings have been
considered in detail in \cite{Lerche:1998nx} and it is a useful
check on our formalism\footnote{and on the results of
\cite{Lerche:1998nx} as well, some of which have been corrected in
the erratum in \cite{Lerche:1998nx}.} to rederive their results.
For $d=2$ (as for $d=1$) the modular integrals can be evaluated
thanks to the summation identities of Appendix \ref{latsub} and the
explicit modular integral \eqref{I2} for the unshifted lattice.
For the $(0,16)$ gauge couplings  \eqref{d4}, \eqref{d22} in a
given $SO(8)$, we can use the identity \eqref{mi4} along with the
explicit result \eqref{I2} to obtain
\begin{subequations}
\begin{equation} \label{d42} \Delta_{\Tr F_{\d}^4}
 =I_2(T,U)-I_2(T/2,U)=-\log \frac{2|\eta(T)|^4}{|\eta(T/2)|^4} \ ,
\end{equation}
\begin{equation}
\label{d222} \Delta_{\Tr F_{\d}^2 \Tr F_{\d}^2 }
=I_2(T/2,U)-\frac12 I_2(T,U)
 = - \frac{1}{2}\log \left[ \frac{2 \pi e^{1-\gamma_E}}{3\sqrt{3}}
  \frac{ T_2 U_2|\eta(T/2)|^8|\eta(U)|^4 }{|\eta(T)|^4 }
  \right] \ .
\end{equation}
\end{subequations}
Under the duality, $T,U$ map to the complex structures $U_{\rm
F}$, $U_{\rm B}$ of the fiber and the base respectively, so that
\eqref{d42}, \eqref{d222}  appear to give a tree-level result
together with an infinite series of D-instanton corrections of
classical action $S_{\rm cl}=N U_{\rm F}$. Such effects have been
discussed in a related context in \cite{Gutperle:1999xu}.

For the couplings \eqref{d22p} between the four different $SO(8)$
factors we have 3 pairs of possibilities, for  which we use the
summation identities \eqref{allsum2}, \eqref{allsum3},
\eqref{allsum4} respectively, yielding after some algebra
\begin{subequations}
\begin{eqnarray}
\Delta_{01}&=&\Delta_{23}=I_2(T/2,U/2)-I_2(T/2,U)=
\log\frac{2|\eta(U)|^4}{|\eta(U/2)|^4} \ , \\
\Delta_{02}&=& \Delta_{13} = I_2(T/2,2U)-I_2(T/2,U)=
\log\frac{|\eta(U)|^4}{2|\eta(2U)|^4} \ , \\
\Delta_{03}&=&\Delta_{12} =I_2(T/2,(U+1)/2)-I_2(T/2,U)=
\log\frac{2|\eta(U)|^4}{|\eta(\frac{U+1}{2})|^4} \ .
\end{eqnarray}
\end{subequations}
In that case, the F-theory couplings arise at one-loop only
from the point of view of the IIB orientifold perturbative description.
This is consistent with the fact that gauge bosons from different
branes have to be inserted on opposite sides of the cylinder
in a one-loop computation.
For the gravitational couplings \eqref{r4}, we find instead
\begin{equation} \label{dr42}
\Delta_{\Tr R^4}=4\Delta_{(\Tr R^2)^2}=\frac{1}{16}I_2(T/2,U)
 =-\frac{1}{16}  \log \left[
\frac{4 \pi e^{1-\gamma_E}}{3\sqrt{3}} T_2 U_2
|\eta(T/2)|^4 |\eta(U)|^8
  \right] \ ,
\end{equation}
which exhibits an infinite series of D-instanton effects from
expanding the Dedekind function.

\subsection{M theory on $K_3$ and enhanced gauge symmetry \label{instm}}
We now turn to the heterotic string compactified on $T^3$ at the
$SU(2)^{16}$ point. One dual description is provided by type IIA
on a $T^3/\Zint_2$ orientifold, which is similar to the two
previous cases. We are however more interested in the M-theory
description, which involves compactification on $K_3$ with $A_1$
conical singularities (see Section \ref{modm}). Each of the 8
fixed points of the $T^3/\Zint_2$ orientifold has thus split into
two distinct fixed points of the $T^4/\Zint_2$ M-theory orbifold.
Using the duality map \eqref{dualm}, we find that the $F^4$
couplings \eqref{d4}, \eqref{d22}, \eqref{d22p} translate into
\begin{subequations}
\begin{equation}
\label{d4m} \Delta_{\Tr F_{\d}^4}
 = \frac{\lm^3}{12\sqrt{V_{K_3}}} \int_{\F_2^-}
\frac{d^2\tau}{\tau_2^2}  \left( 2 \Zar{0}{0} -
\Zar{0}{\d} \right) (\gamma,\beta) \ ,
\end{equation}
\begin{equation}
\label{d22m} \Delta_{(\Tr F_{\d}^2)^2} =
\frac{\lm^3}{12\sqrt{V_{K_3}}} \int_{\F_2^-}
\frac{d^2\tau}{\tau_2^2}  \Zar{0}{\hd} (\gamma,\beta) \ ,
\end{equation}
\begin{equation}
\label{d22mp} \Delta_{\Tr F_{\d_1}^2 \Tr F_{\d_2}^2 } =
\frac{\lm^3}{12\sqrt{V_{K_3}}} \int_{\F_2^-}
\frac{d^2\tau}{\tau_2^2} \left( 2Z\ar{0}{\hd} - 3 Z\ar{00}{\d 1}
\right) (\gamma,\beta) \ ,
\end{equation}
\end{subequations}
where $(\gamma,\beta)$ encode the shape of the orbifold
$T^4/\Zint_2$. The 3-digit numbers $\d$ label one of the 8 copies
of the gauge group $SO(4)=SU(2)\times SU(2)$. We can rewrite these
results in a more appealing fashion by using the representation
\eqref{I3} of the modular integrals in terms of $SO(3,3,\Zint)$
Eisenstein series in the spinor representation along with the
identity \eqref{mi4}. For example, the $\Tr (F_\d)^4$ coupling
\eqref{d4m} can be rewritten as
\begin{equation}
\label{d4me} \Delta_{\Tr F_{\d}^4}
 = \frac{\lm^3}{2\pi\sqrt{V_{K_3}}}
\left[ \eis{SO(3,3,\Zint)}{S}{s=1}(\gamma,\beta)
- \sqrt{2} \eis{SO(3,3,\Zint)}{S}{s=1}(\gamma/2,\beta/2) \right] \ ,
\end{equation}
where the $SO(3,3,\Zint)$ Eisenstein series is defined by
\cite{Obers:1999um}:
\begin{equation}
\eis{SO(3,3,\Zint)}{S}{s=1}(\gamma,\beta)
=\hat{\sum_{m^{1,2,3},n\in \Zint}} \frac{\sqrt{\det\gamma}}
{(m^i+\beta^i n)\gamma_{ij}(m^j+\beta^j n)+(\det\gamma) n^2}
\end{equation}
with $\beta^i=\epsilon^{ijk}\beta_{jk}/2$, and the hat restricts
the sum to non-zero integers. Using \eqref{bg}, we can re-express
this in terms of the metric $G$ of the type IIA orbifold
\begin{equation}
\eis{SO(3,3,\Zint)}{S}{s=1}(\gamma,\beta)
=\hat{\sum_{m^r\in \Zint}} \frac{\sqrt{V_{K_3}}}{m^r G_{rs} m^s}
=\sqrt{V_{K_3}} \eis{Sl(4,\Zint)}{4}{s=1}(G) \ ,
\end{equation}
where $m_r$ can be thought of as momenta along $T^4$.
We can now rewrite \eqref{d4me} in terms of Eisenstein series
for a congruence 2 subgroup of $Sl(4,\Zint)$,
\begin{equation}
\label{d4mf} \Delta_{\Tr F_{\d}^4}
 = \frac{\lm^3}{2\pi}
\left[ \hat{\sum_{m^r\in \Zint}} \frac{1}{(m^r G_{rs} m^s)}
-4 \hat{\sum_{\substack{ m^{2,3,4}\in 2\Zint\\m^1\in \Zint}}}
\frac{1}{(m^r G_{rs} m^s)} \right] \ .
\end{equation}
The fact that the direction 1 is singled out should not come
as a surprise, since the 16 orbifold fixed points originate
from the 8 orientifold points which have split along direction 1.
The 16 fixed points should however appear on the same footing
from the M-theory point of view. This is indeed so, since, upon
decomposing the $SO(4)$ gauge field $F_{\d}=F_{\d0}\otimes 1 +
1\otimes F_{\d 1}$
into its $SU(2)\times SU(2)$ components and using the identity
$\Tr F^4 = (\Tr F^2)^2$ for $SU(2)$ gauge fields, we have
\begin{subequations}
\begin{equation}
\Tr F^4=\Tr F_{\d0}^4 + \Tr F_{\d 1}^4 + 6 \Tr F_{\d0}^2 \Tr F_{\d 1}^2 \ ,
\end{equation}
\begin{equation}
(\Tr F^2)^2 = \Tr F_{\d0}^4 + \Tr F_{\d 1}^4 + 2 \Tr F_{\d0}^2 \Tr F_{\d 1}^2
\ .
\end{equation}
\end{subequations}
The $SO(4)$ gauge couplings \eqref{d4mf} can thus be rewritten as
$SU(2)$ gauge couplings
\begin{subequations}
\label{d4mf2}
\begin{equation}
\Delta_{\Tr F_{\d0}^4}=\Delta_{\Tr F_{\d1}^4}
 = \frac{\lm^3}{4\pi}
\hat{\sum_{m^r\in \Zint}} \frac{1}{(m^r G_{rs} m^s)} \ ,
\end{equation}
\begin{equation}
\Delta_{\Tr F_{\d0}^2\Tr F_{\d1}^2}= \frac{\lm^3}{2\pi} \left[ 5
\hat{\sum_{m^r\in \Zint}} \frac{1}{(m^r G_{rs} m^s)} - 16
\hat{\sum_{\substack{ m^{2,3,4}\in 2\Zint\\m^1\in \Zint}}}
\frac{1}{(m^r G_{rs} m^s)}\right] \ .
\end{equation}
\end{subequations}
The $\Tr F_{\d0}^4$ now makes no reference to any particular direction
as it should, while the $\Tr  F_{\d0}^2\Tr F_{\d1}^2$ singles out
the direction 1 along which the two gauge fields are separated.

The results \eqref{d4mf2} are given
 exactly at first order  in $\lm^2/\sqrt{V_{K_3}}$,
which is the natural expansion parameter on the M-theory side problem.
They cannot however be obtained from eleven-dimensional supergravity
in perturbation theory due to the conical singularity, and it is
necessary to include the M2-brane  in order to provide the
$SU(2)$ degrees of freedom. It would be very interesting to devise
a perturbative approach in this situation, perhaps along the lines
of \cite{Witten:1996mz}, in order to recover the result \eqref{d4mf}.
We also note that the $F^4$ couplings that we have computed in
M-theory on $T^4/\Zint_2$ also give the $F^4$ couplings in type IIA on
$T^4/\Zint_2$ in the absence of B-flux on the vanishing cycles, where the
conformal field theory is singular. Surprisingly, they are finite. They
should presumably correspond to the finite part of the $F^4$ amplitude
when the singularity has been subtracted, and it would be interesting to
analyze the behaviour of the amplitude when the B-flux is perturbed away
from zero.    

\subsection{Type IIA on $K_3 \times S_1$, IIB on $K_3$
and D-instantons \label{instii}}

For $d>4$, the dual description of the heterotic string compactified
on $T^d$ at the special $U(1)^{16}$ point now allows for
non-perturbative effects. In particular, for $d=5$, the type IIA string
theory compactified on $K_3\times S_1$ has instanton configurations
coming from even D-branes whose Euclidean world-volume is wrapped
on even-cycles of $K_3$ times the circle $S_1$. In the type IIB picture,
instanton configurations exist already in 6 dimensions, as odd Euclidean
D-branes wrapped on even cycles of $K_3$. These effects were first
computed in \cite{Antoniadis:1997zt}, and here we want to get
a more quantitative understanding of them.

From the duality relation $R_{\rm H}/\lh= R_{\rm
A}/(\giis\lii)=1/\giibs$, we see that the weak coupling regime on
the type IIA or IIB side corresponds to the decompactification
limit $R_{\rm H}\gg \lh$ on the heterotic side. In this limit, the
heterotic result exhibits a series of world-sheet instanton
contributions which will be interpreted as D-instanton effects on
the type II side. For simplicity, we will focus on the
$\Tr(F_\d)^4$ couplings in \eqref{d4p}, given by a modular
integral of an unshifted partition function,
\begin{equation}
\label{del5} \Delta_{5D} = \lh^3 \int_{\F}
\frac{d^2\tau}{\tau_2^2} Z_{5,5}(g/\lh^2,b,\rh/\lh,w) \ ,
\end{equation}
where we dropped the numerical factor, and denoted by $w$ the Wilson
lines of the six-dimensional $(4,4)$ gauge fields around the extra
circle. We will comment on the effects of shifts at the end.

In order to determine the large $\rh$ behaviour of \eqref{del5},
it is convenient to adopt a Lagrangian representation for the
$S^1$ part and a Hamiltonian representation for the $T^4$ part:
\begin{gather}
\label{del5lh} \Delta_{5D}=\lh^{2} \rh \int_{\cal F}
\frac{d^{2}\tau}{\tau_2^2} \sum_{p,q} \sum_{m_i,n^i} \exp\left( -
\pi \frac {\rh^2|p-\tau q|^2 }{\lh^2 \tau_2} + 2\pi i p~ w_i n^i
\right) \tau_2^2 q^{\frac{p_L^2}{2}} \bar q^{\frac{p_R^2}{2}} \ ,
\end{gather}
where $m_i,n^i$ denote the momenta and windings on $T^4$.
We apply the standard orbit decomposition method on the integers $(p,q)$,
trading the sum over $Sl(2,\Zint)$ images of $(p,q)$ for
a sum over images of the fundamental domain ${\cal F}$ \cite{Dixon:1991pc}
(see \cite{Kiritsis:1997hf,Obers:1999um} for relevant formulae).
The zero orbit gives back the six-dimensional result \eqref{trial5}
up to a volume factor, and reproduces the
tree-level type II contribution in 5 dimensions:
\begin{gather}
\label{del5z} \Delta_{5D}^{\rm zero}=\lh^{2} \rh \int_{\cal F}
\frac{d^{2}\tau}{\tau_2^2} Z_{4,4}(g/\lh^2,b) = R_{\rm A}
\Delta_{\rm IIA}^{\rm tree} \ .
\end{gather}
The degenerate orbit on the other hand, with representatives
$(p,0)$, can be unfolded onto the strip $|\tau_1|<1/2$. The $\tau_1$
integral then imposes the level matching condition $p_L^2-p_R^2=2m_i n^i=0$,
and the $\tau_2$ integral can be carried out in terms of Bessel
functions to give
\begin{equation}
\label{del5d} \Delta_{5D}^{\rm deg}=2 \lh \rh^2 \sum_{p\neq 0}
\sum_{(m_i,n^i)\neq 0}  \delta(m_i n^i) \frac{|p|}{\sqrt{m^t
M_{4,4} m}} K_1\left( 2\pi \frac{\rh}{\lh} |p| \sqrt{m^t M_{4,4}
m} \right) e^{2\pi i p w_i n^i}\ ,
\end{equation}
up to a divergent contribution $\pi^2 \rh^3 \Gamma(-1)/3$,
coming from the origin of the $(4,4)$ lattice, which we assume to
be regularized. Here $m^t M_{4,4} m= p_L^2+p_R^2$ with
$m=(m^i,n_i)$. It is straightforward to translate this result to
the type IIA side,
\begin{align}
\label{del5d2} \Delta_{5D}^{\rm deg}=2 \giis \lii R_{\rm A}^2
\sum_{p\neq 0} \sum_{(m_i,n^i)\neq 0} & \delta(m_i n^i) \\
\cdot & \frac{|p|}{\sqrt{m^t M_{4,4} m}} K_1\left( 2\pi \frac{R_{\rm
A}}{\giis\lii}
 |p| \sqrt{m^t M_{4,4} m} \right) e^{2\pi i p w_i n^i}\ ,\nonumber
\end{align}
where $M_{4,4}$ is now the mass matrix \eqref{mass420} of D-brane
states wrapped on the untwisted cycles of $T^4/\Zint_2$.
Given the asymptotic behaviour $K_1(x)\sim \sqrt{\frac{\pi}{2x}}e^{-x}$,
we see that this is a sum of order $e^{-1/g_s}$ non-perturbative effects
corresponding to $N=p r$ Euclidean (anti) D-branes wrapped on $S_1$ times
a cycle of homology charges $(m_i,n^i)/r$ on $T^4$, where $r$ is the
greatest common divisor of $(m_i,n^i)$.

It is worth pointing out a number of peculiarities of the 
result \eqref{del5d2}.
First, due to the absence of a holomorphic insertion in
\eqref{del5}, all instanton effects are due to {\it untwisted}
D-branes wrapped along even cycles of $K_3$, even though we are
discussing $F^4$ couplings between fields located on the fixed
points of the orbifold. This is in contrast to the result in
four-derivative scalar couplings \cite{Antoniadis:1997zt}, where a
contribution from the whole Hagedorn density of BPS states was found.
This is an important simplification due to our choice of the
orbifold point in the $K_3$ moduli space. Second, the integration
measure corresponding to a given number of D-branes $N$ is easily
seen to be $\sum_{r|N} (1/r^2)$, where $r$ runs over the divisors
of $N$, just as in the case of D-instanton effects in theories
with 32 supersymmetries \cite{Green:1997tv,Green:1998tn}. This is
an unexpected result, since the bulk contribution to the 
index for the quantum mechanics
with 8 unbroken symmetries is $1/N^2$ instead \cite{Moore:1998et},
which did arise in four-derivative scalar couplings at the enhanced symmetry
point \cite{Antoniadis:1997zt,Green:1998yf}. Finally, it is clear that
the above analysis goes through in the case with shifts on the
lattice, since those only affect the momenta and windings on
$T^4$. They translate into corresponding shifts on the lattice of
D-instantons contributing to $\Tr F^4$.

The situation from the T-dual type IIB point of view is also
interesting. From the mapping \eqref{dualiib}, we see that the one-loop
heterotic $F^4$ coupling in 5 dimensions translates into
\begin{equation}
\label{d5b} \Delta_{5D}=\left(\frac{\lp^2}{R_{\rm B}}\right)^{3}
\int_{\F} \frac{d^2\tau}{\tau_2^2} Z_{5,5} \ ,
\end{equation}
where now $Z_{5,5}$ depends on the $K_3$ untwisted moduli and on
the six-dimensional string coupling, but not on the size of the
circle $S^1$ in six-dimensional Planck units. We can therefore
simply take the limit $R_{\rm B}\to \infty$ to recover a
six-dimensional amplitude. The powers of $R_{\rm B}$ in \eqref{d5b}
are precisely such as to yield a finite $t_{12} H^4$ coupling in 6
dimensions, where $H$ is one of the 16 anti-self-dual three-form
field strengths arising from the twisted sectors of type IIB
compactified on $T^4/\Zint_2$, and $t_{12}$ is a 12-index tensor
constructed from $t_8$. We therefore get
\begin{equation}
\label{d6b} \Delta_{H^4}^{\rm IIB}= \lp^6 \int_{\F}
\frac{d^2\tau}{\tau_2^2} Z_{5,5}
\end{equation}
which is the exact non-perturbative coupling of four self-dual
twisted three-forms, invariant under the $SO(5,5,\Zint)$ subgroup
of the U-duality group $SO(5,21,\Zint)$ left unbroken by the
choice of the external legs. The above analysis of the heterotic
decompactification limit still holds, and yields the tree-level
and D-instanton contributions to this amplitude,
\begin{align}
\Delta_{H^4}^{\rm IIB}=&\frac{\gii^2 \lii^{10}}{V_{K_3}}
\int_{\cal F} \frac{d^{2}\tau}{\tau_2^2} Z_{4,4} +2
\left(\frac{\gii^2 \lii^{10}}{V_{K_3}} \right)^{3/2} \cdot \nn
\\&\cdot \sum_{p\neq 0} \hat{\sum_{m_i,n^i}} \delta(m_i n^i)
\frac{ \sqrt{m^t M_{4,4} m} }{|p|} K_1\left( 2\pi
\frac{V_{K_3}^{1/2}}{\gii \lii^2}
 |p| \sqrt{m^t M_{4,4} m} \right)
e^{2\pi i p w_i n^i}
\end{align}
which exhibits non-perturbative contributions from odd
D-branes wrapped on even untwisted cycles of $K_3$. In
particular, the ten-dimensional decompactification limit
$V_{K_3}\gg \lii^4$ reproduces the $R^4$ couplings in type IIB,
as demonstrated in \cite{Antoniadis:1997zt}.

\subsection{Type II on $K_3\times T^2$ and NS5-brane corrections\label{inst4}}
Finally, we would like to discuss the four-dimensional case, which on the
type II side receives corrections from NS5-branes wrapped on $K_3\times
T^2$. Similar corrections could also in principle arise on the heterotic
side from 5-branes wrapped on $T^6$, but they do not affect four-gauge-boson
couplings from the right-moving sector according to our conjecture.
From the duality map $T_{\rm H}=S_{\iia}=S_{\iib}$, the weak coupling
regime on the type II side again corresponds to the limit where the
heterotic $T^2$ decompactifies.

The study of the decompactification limit proceeds as in
\eqref{del5lh} by performing an orbit decomposition on the
integers running in the Lagrangian representation of the $T^2$
lattice, and the zero orbit and degenerate orbit reproduce the
tree-level and D-instanton contributions on the type II side. The
novelty in that case is that there is a third orbit, namely the
non-degenerate orbit, which contributes as well. The
integral on $\tau_1$ is Gaussian, and the subsequent integral
along $\tau_2$ is again given by a Bessel function. Before
carrying out this integration, it is more enlightening to
determine the saddle point, which controls the instanton effects
at leading order. The saddle point equations are easily found to
be
\begin{subequations}
\begin{eqnarray}
q^I g_{IJ} (p^J-\tau_1 q^J)+ i \tau_2 m_i n^i&=&0 \ , \\
-(p^I-\tau_1 q^I)g_{IJ}(p^J-\tau_1 q^J)+\tau_2^2(q^I g_{IJ} q^J
+ m^t M_{4,4}m)&=&0 \ ,
\end{eqnarray}
\end{subequations}
where $p^I$ and $q^I$ are the integers running in the $T^2$ lattice
partition function, and should be summed over $Sl(2,\Zint)$ orbits
such that $p^1 q^2-p^2 q^1\neq 0$
only. $g_{IJ}$ is the metric on $T^2$ in heterotic units.
The solution of these equations is given by
\begin{subequations}
\begin{eqnarray}
\tau_1&=&\frac{pq}{q^2} + i
\frac{m_i n^i}{q^2} \sqrt{
\frac{p^2 q^2 - (pq)^2 }{(q^2)^2+ q^2 m^t M_{4,4} m + (m_i n^i)^2} } \\
\tau_2&=&\sqrt{\frac{p^2 q^2 - (pq)^2 }{(q^2)^2+ q^2 m^t M_{4,4} m
+ (m_i n^i)^2 \ ,
}}
\end{eqnarray}
\end{subequations}
 where contractions with $g_{IJ}$ are understood,
and corresponds to a classical action
\begin{align} S_{\rm cl}= 2\pi &
\sqrt{ ( p^2 q^2 - (pq)^2)  \left( 1 + \frac{m^t M_{4,4}
m}{q^2} + \frac{(m_i n^i)^2}{(q^2)^2} \right) } \nn \\ &+ 2\pi i
\frac{ (pq) (m_i n^i)}{q^2} + 2\pi i pBq \ .
\end{align}
Reinstating the $\lh$ dependence and mapping to dual type IIA variables
using \eqref{hetiia}, the real part of the classical action
\begin{equation}
\Re S_{\rm cl}=2\pi \sqrt{ \frac{p^2 q^2- (pq)^2}{(q^2)^2}
\left( \frac{(q^2)^2}{\giis^4 \lii^4}
 + \frac{q^2 m^t M_{4,4} m }{\giis^2 \lii^2} + (m_i n^i)^2 \right) }
\end{equation}
scales as $1/\giis^2$. The corresponding non-perturbative
effects should therefore be interpreted as coming from
$N=|p^1 q^2-p^2 q^1|$ NS5-branes wrapped on $K_3\times T^2$,
and bound to D-brane states wrapped on an even cycle of $K_3$
times a circle on $T^2$ determined by the integers $q^1,q^2$.
The result of the $\tau$ integration thus gives
\begin{equation}
\Delta_{4D}^{\rm n.d.}=4 \lh^{4} \sum_{p^i,q^i} \sum_{m_i,n^i}
\left(\frac{(q^2)^2+q^2 m^t M_{4,4}
m+(m_in^i)^2}{p^2q^2-(pq)^2}\right)^{3/4} K_{3/2} \left( \Re
S_{\rm cl} \right) e^{i \Im S_{\rm cl}} \ .
\end{equation}
In particular, we may look at the contribution of pure
NS5-brane instantons, corresponding to $m_i=n^i=0$.
Choosing the orbit representatives as
\begin{equation}
\begin{pmatrix} q^1 & p^1 \\ q^2 & p^2 \end{pmatrix}
=
\begin{pmatrix} k    & j \\ 0 & p \end{pmatrix}\ ,
0\leq j < k\ , \quad p\neq 0\ ,
\end{equation}
and using the exact expression for the Bessel function
\begin{equation}
\label{k32} K_{3/2}(x)=\sqrt{\pi/2x}\left(1+1/x \right) e^{-x}
\end{equation}
we obtain
\begin{equation}
\Delta_{4D}^{\rm NS5}= 2 (\giis \lii)^4 U_2 \sum_N \mu (N)  \left(
N+\frac{1}{2\pi S_2} \right) e^{-2\pi N S_2} \left( e^{2\pi i N
S_1}+e^{-2\pi i N S_1} \right) \ ,
\end{equation}
where we used the type II variable $S=a+iV_{K_3} V_{T^2}/(\gii^2 \lii^6)$
and extracted the instanton measure
\begin{equation}
\label{mu5} \mu (N)=\sum_{r|N} \frac{1}{r^3} \sp (\mbox{NS5-brane
on}\;\,K_3 \times T^2) \ .
\end{equation}
This result gives a prediction for the index (or
rather the bulk contribution thereto) of the world-volume
theory of the type II NS5-brane wrapped on $K_3\times T^2$. It
is a challenging problem to try and derive this
result from first principles. It is
also remarkable that, in virtue of \eqref{k32} and in contrast to
D-instantons, the NS5-instantons contributions do not seem to
receive any perturbative subcorrections beyond one-loop.

It is interesting to compare this result to the corresponding
index of the heterotic 5-brane wrapped on $T^6$, which can be extracted
from the non-perturbative $R^2$ couplings in the heterotic string
compactified on $T^6$ \cite{Hammou:1999in,Kiritsis:1999ss}.
Those can be computed by duality from
the one-loop
exact $R^2$ couplings in type  II on $K_3\times T^2$ \cite{Harvey:1996ir},
and read
\begin{subequations}
\begin{align}
 \Delta_{R^2} =\hat{\cal E}^{Sl(2,\Zint)}_{\irrep{2};s=1}
  & = -\pi \log (S_2 |
\eta(S) |^4) \\ & = \frac{\pi^2}{3} S_2 + 2 \pi \sqrt{S_2}
\sum_N \mu(N)e^{-2\pi N S_2} \left( e^{2\pi i N S_1}+e^{-2\pi i N
S_1} \right) \ .
\end{align}
\end{subequations}
The summation measure turns out to be different from \eqref{mu5} and
given instead by
\begin{equation}
\mu (N)=\sum_{r|N} \frac{1}{r} \sp (\mbox{Het 5-brane on}\;\, T^6) \ .
\end{equation}
It is also worthwhile to notice that there are no subleading corrections
around the instanton  in  the heterotic 5-brane case, whereas, by
 virtue of \eqref{k32}, these
corrections occur at first order only in the type II NS5-brane on
$K_3\times T^2$. This is in contrast to
D-instantons, for which the
saddle point approximation to the Bessel function $K_{1}$ is not exact.
 It would be interesting to have a deeper understanding
of these non-renormalization properties, possibly  using the CFT
description of the 5-brane \cite{Callan:1991at}.

\newpage
\centerline{\bf \Large Appendices}

\appendix

\section{Shifted partition functions and lattice integrals \label{lat}}

\subsection{Hamiltonian and Lagrangian representation \label{lath}}
As discussed in Section \ref{hetao}, the compactification on a torus
with half-integer Wilson lines \eqref{hety} is most conveniently
described in terms of shifted lattice sums, which in the
Hamiltonian representation read
\begin{equation}
\label{zhgdef} Z_{d,d}\ar{h^i}{g^i}(g,b,\tau)=\tau_2^{d/2}
\sum_{m_i,n^i\in\Zint} (-)^{m_i g^i} q^{\frac12 p_L^2} \bar
q^{\frac12 p_R^2}\ .
\end{equation}
The left-moving and right moving momenta $p_L,p_R$ are given by
\begin{equation}
\label{prl} p_{\substack{L\\R}}^i =n^i+\frac{h^i}{2} \pm g^{ij}
\left[m_i+ B_{ij} \left(n^j + \frac{h^j}{2} \right) \right]\ ,
\end{equation}
and the integers $h^i,g_i$ are defined modulo 2, and when
non-zero, break the T-duality $O(d,d,\Zint)$ to a finite index
subgroup. Modular invariance on the other hand is manifest in the
Lagrangian representation, obtained after Poisson resumming on the
momenta $m_i$:
\begin{equation}
\label{zhglag} Z_{d,d}\ar{h^i}{g^i}(g,b,\tau)=V \sum_{
\substack{m^i\in\Zint+g^i/2\\n^i\in\Zint+h^i/2} }
\exp\left(-\frac{\pi}{\tau_2} \left(m^i-\tau
n^i\right)g_{ij}\left(m^i-\bar\tau n^i\right) +2\pi i m^i B_{ij}
n^j \right) \ .
\end{equation}
In particular, insertions of left-moving and right-moving momenta
in the Hamiltonian representation translate into
\begin{equation}
\label{inslag} p_L^i \rightarrow - \frac{m^i+n^i\bar \tau}{i
\tau_2}\ ,\quad p_R^i \rightarrow \frac{m^i+n^i\tau}{i \tau_2} \ ,
\end{equation}
where the $m^i$ and $n^i$ are integers shifted by $g^i/2$ and
$h^i/2$ respectively. This translation is up to contractions which
are easily fixed by demanding modular invariance. In particular,
under modular transformations of $\tau$, $p_L$ and $p_R$ have
modular weight (1,0) and (0,1) respectively.

When $h^i$ or $g^i$ is non-zero, the shifted blocks \eqref{zhglag}
are not modular invariant. Instead, they transform among
themselves as
\begin{subequations}
\begin{eqnarray}
\label{zhgst} T &:&\qquad Z_{d,d}\ar{h^i}{g^i}
(\tau+1)=Z_{d,d}\ar{h^i}{g^i+h^i}(\tau)\ ,\\ S &:&\qquad
Z_{d,d}\ar{h^i}{g^i}\left(-\frac1{\tau} \right)=
Z_{d,d}\ar{g^i}{h^i} (\tau) \ ,
\end{eqnarray}
\end{subequations}
so that like T-duality, modular invariance is broken to a finite
index subgroup, namely the subgroup of $Sl(2,\Zint)$ leaving all
$(h^i,g^i)$ invariant modulo 2. It will be quite useful to have a
precise understanding of these subgroups, to which we now turn.

\subsection{Congruence 2 subgroups of $Sl(2,\Zint)$ \label{latsub}}
Under the modular group $Sl(2,\Zint)$, the characteristics $(h,g)$
transform as a doublet. The subgroup of $Sl(2,\Zint)$ leaving
$(h,g)$ invariant modulo 2 is easily found to be
\begin{subequations}
\begin{eqnarray}
\label{g2+} \ar{h}{g}&=&\ar{1}{0}\ :\qquad
   \Gamma_2^+:= \begin{pmatrix} 1 & 0 \\ * & 1 \end{pmatrix}
\ ,\quad \left\{ \begin{matrix} T^2:\tau\to\tau+2 \\ STS:\tau\to
\tau/(1-\tau) \end{matrix} \right. \ ,
\\
\label{g2-} \ar{h}{g}&=&\ar{0}{1}\ :\qquad \Gamma_2^-:=
    \begin{pmatrix} 1 & * \\ 0 & 1\end{pmatrix}
\ ,\quad \left\{ \begin{matrix} T:\tau\to\tau+1 \\ ST^2S:\tau\to
\tau/(1-2\tau) \end{matrix} \right. \ ,
\\
\label{g20} \ar{h}{g}&=&\ar{1}{1}\ :\qquad \Gamma_2^0:=
  \begin{pmatrix} 1 & 0 \\ 0 & 1\end{pmatrix}
  \mbox{ or }
  \begin{pmatrix} 0 & 1 \\ 1 & 0 \end{pmatrix}
\ ,\quad \left\{ \begin{matrix} T^2:\tau\to\tau+2 \\ S:\tau\to
-1/\tau \end{matrix} \right. \ ,
\end{eqnarray}
\end{subequations}
where we represented the subgroups by the value of the allowed
matrices modulo 2 (where $*$ stands for 0 or 1), and listed their
generators. These three subgroups are of index 3 in $Sl(2,\Zint)$,
and correspond to the invariance groups (modulo phases and
weights) of $\th_4,\th_2,\th_3$ respectively. Equivalently, they
are the invariance subgroups of $Z(\tau/2),Z(2\tau),Z((\tau+1)/2)$
respectively, where $Z$ is an $Sl(2,\Zint)$  modular form. The
intersection of any two of these subgroups gives the index 6
subgroup of $Sl(2,\Zint)$
\begin{equation}
\label{g2} \Gamma_2:= \begin{pmatrix} 1 & 0 \\ 0 & 1 \end{pmatrix}
\ ,\quad \left\{ \begin{matrix} T^2:\tau\to\tau+2 ,\\
ST^2S:\tau\to \tau/(1-2\tau) \end{matrix} \right. \ .
\end{equation}
Therefore, for several non-vanishing shifts $(h^i,g^i)$, the
unbroken group is either $\Gamma_2^{-,0,+}$ if all the $(h^i,g^i)$
are the same, or  $\Gamma_2$ if they are different. The lattice
sums \eqref{zhglag} hence either form a length-3 orbit in the
first case, or a length-6 orbit in the second.

The fundamental domains ${\cal F}_2^{+,0,-}$ of the
upper-half-plane for the groups $\Gamma_2^{-,0,+}$ are three-fold
and six-fold covers respectively of the fundamental domain ${\cal
F}$ of $Sl(2,\Zint)$. Integral over these fundamental domains can
be converted into each other at the expense of introducing
appropriate orbits. In particular, we have, for a $\Gamma_2^-$
modular invariant function $\Phi$,
\begin{equation}
\label{fi1} \int_{{\cal F}_2^-} \frac{d^2\tau}{\tau_2^2}
\Phi(\tau)
=
\int_{{\cal F}} \frac{d^2\tau}{\tau_2^2} \left[ \Phi(\tau)
+\Phi\left(-\frac{1}{\tau}\right)+\Phi\left(-\frac{1}{\tau+1}\right)
\right]
\end{equation}
and for an $Sl(2,\Zint)$ modular invariant function $Z$,
\begin{align}
\label{fi2} \int_{{\cal F}} \frac{d^2\tau}{\tau_2^2}& \left[
Z(2\tau)
+Z\left(\frac{\tau}{2}\right)+Z\left(\frac{\tau+1}{2}\right)
\right] = \nn  \\ &=\int_{{\cal F}_2^-} \frac{d^2\tau}{\tau_2^2}
Z(2\tau) = \int_{{\cal F}_2^+} \frac{d^2\tau}{\tau_2^2}
Z\left(\frac{\tau}{2}\right) = \int_{{\cal F}_2^0}
\frac{d^2\tau}{\tau_2^2} Z\left(\frac{\tau+1}{2}\right) \ .
\end{align}
Moreover, by changing integration variables to $\rho=2\tau$, this
can yet be rewritten as
\begin{equation}
\label{fi3} \int_{{\cal F}_2^-} \frac{d^2\tau}{\tau_2^2} Z(2\tau)
=\int_{{\cal F}_2^+} \frac{d^2\rho}{\rho_2^2} Z(\rho) =3\int_{\cal
F} \frac{d^2\tau}{\tau_2^2} Z(2\tau) \ .
\end{equation}

\subsection{Summation identities \label{latsum}}
Since the string world-sheet theory is modular invariant, the
shifted sums \eqref{zhgdef} have to appear in modular invariant combinations.
These combinations amount to projecting the original unshifted
partition function $Z_{d,d}\ar{0}{0}$ to even momenta or add
half-integer winding sectors. As a result, they can be re-expressed
as unshifted partition functions of tori with different moduli. In
particular,
\begin{equation}
\label{allsum} \sum_{\d,\d'} Z_{d,d}\ar{\d'}{\d}(g,b;\tau)= 2^d
Z_{d,d}(g/4,b/4;\tau) \ .
\end{equation}
In particular, for $d=1$ we have
\begin{equation}
\label{allsum1}
\frac{1}{2}
\left(Z_{1,1}\ar{0}{0}(R)+Z_{1,1}\ar{0}{1}(R)
+Z_{1,1}\ar{1}{0}(R)+Z_{1,1}\ar{1}{1}(R) \right)
=Z_{1,1}(R/2) \ .
\end{equation}
For $d=2$, we will also need the following identities
(see for instance \cite{Gregori:1997hi}):
\begin{subequations}
\begin{equation}
\label{allsum2}
\frac{1}{2} \left(
Z\ar{00}{00}+Z\ar{00}{10}+Z\ar{10}{00}+Z\ar{10}{10} \right)
=Z(T/2,2U) \ ,
\end{equation}
\begin{equation}
\label{allsum3}
\frac{1}{2} \left(
Z\ar{00}{00}+Z\ar{00}{01}+Z\ar{01}{00}+Z\ar{01}{01} \right)
=Z(T/2,U/2) \ ,
\end{equation}
\begin{equation}
\label{allsum4}
\frac{1}{2} \left(
Z\ar{00}{00}+Z\ar{00}{11}+Z\ar{11}{00}+Z\ar{11}{11} \right)
=Z (T/2,(U+1)/2) \ ,
\end{equation}
\begin{align}
\label{allsum5} \frac{1}{2} \left(
Z\ar{01}{10}\right. &\left.+Z\ar{10}{01} +Z\ar{01}{11}+Z\ar{10}{11}
+Z\ar{11}{01}+Z\ar{11}{10} \right) =\\
&2Z(T/4,U)-Z(T/2,2U)-Z(T/2,U/2)-Z(T/2,(U+1)/2)+Z(T,U) \ .  \nonumber
\end{align}
\end{subequations}
We also note the partial sums, valid for any $d$,
\begin{subequations}
\label{partisum}
\begin{eqnarray}
 Z_{d,d}\ar{0}{\d}(\tau)&=&2^{d/2} Z_{d,d} \left(g/2,b/2;
2\tau\right) \ ,\\ Z_{d,d}\ar{\d}{0}(\tau)&=&2^{d/2} Z_{d,d} \left(g/2,b/2;
\frac{\tau}{2}\right)\ , \\ Z_{d,d}\ar{\d}{\d}(\tau)&=&2^{d/2}
Z_{d,d} \left(g/2,b/2; \frac{\tau+1}{2}\right) \ ,
\end{eqnarray}
\end{subequations}
where the summation over the $d$-digit numbers $\d$ is implicit.
This shows that the three sums in \eqref{partisum} form a length-3
orbit of $Sl(2,\Zint)$.

\subsection{Lattice integral on extended fundamental domain \label{latint}}
We now would like to evaluate modular integrals of the form
\begin{equation}
\label{modint} I_{d,d}[\Phi]= \int_{{\cal F}_2^-}
\frac{d^2\tau}{\tau_2^2} Z_{d,d}\ar{0}{\d}(\tau,\bar \tau) \Phi(\bar \tau)\ ,
\end{equation}
where $\Phi(\tau)$ is an almost holomorphic form invariant under
the index 2 subgroup $\Gamma_2^-$ of $Sl(2,\Zint)$, a typical
example being $\Phi(\tau)=(\alpha E_4 + \beta \et^2)
\th_3^8\th_4^8/\eta^{24}$. The sum over $\d=0\dots 2^{p}-1$ is
implicit, and we shall focus here on $p=d$, even though many of
the results can be extended to the less symmetric case $p<d$. For
$\Phi=1,d=2$, this integral has been computed in
\cite{Mayr:1993mq} and later in \cite{Kiritsis:1998en} by a
different method. For $\Phi\neq 1$ and $d=2$, the basic
observations have been made in \cite{Lerche:1998nx}, and we will
streamline and greatly extend their result to all $d$.

In order to compute this integral, we first convert the shifted
lattice sum $Z\ar{0}{\d}$ into a standard unshifted sum using
\eqref{partisum}, and then change variables to $\rho=2\tau$ as in
\eqref{fi3}. We obtain
\begin{equation}
\label{mi2} I_{d,d}[\Phi]= 2^{d/2} \int_{{\cal F}_2^+}
\frac{d^2\rho}{\rho_2^2} Z(g/2,b/2,\rho)
\Phi\left(\frac{\bar\rho}{2}\right) \ .
\end{equation}
We then unfold the integral on the extended fundamental domain
${\cal F}_2^+$ into an integral on the fundamental domain of
$Sl(2,\Zint)$:
\begin{equation}
\label{mi3} I_{d,d}[\Phi]= 2^{d/2} \int_{{\cal F}}
\frac{d^2\rho}{\rho_2^2} Z(g/2,b/2,\rho) \left[
\Phi\left(\frac{\bar\rho}{2}\right) +
\Phi\left(-\frac{1}{2\bar\rho}\right) +
\Phi\left(\frac{\bar\rho+1}{2}\right) \right]\ .
\end{equation}
Using the definition of the Hecke operator on a $\Gamma_2^-$ modular
form of weight $w$,
\begin{eqnarray}
\label{hg2-}
H_{\Gamma_2^-}\cdot \Phi(\tau)=\frac{1}{2} \left(
\tau^{-w} \Phi \left(-\frac{1}{2\tau}\right)
+ \Phi \left(\frac{\tau}{2}\right)+ \Phi \left(\frac{\tau+1}{2}\right) \right)
\ ,
\end{eqnarray}
we recognize in \eqref{mi3} the action of this operator
on the modular form $\Phi$:
\begin{equation}
\label{mi31} I_{d,d}[\Phi]= 2^{\frac{d}{2}+1} \int_{{\cal F}}
\frac{d^2\tau}{\tau_2^2} Z(g/2,b/2;\tau)~ 
H_{\Gamma_2^-} \cdot \Phi(\bar\tau) \ .
\end{equation}
This operator maps $\Gamma_2^-$ modular forms into $Sl(2,\Zint)$
modular forms and preserves the weight. $H_{\Gamma_2^-}\cdot \Phi$ is
therefore an almost holomorphic form of $Sl(2,\Zint)$ of zero
weight, so that \eqref{mi31} is well defined. We can now use the
standard techniques to express this integral as a sum over zero,
degenerate and non-degenerate orbits. A great simplification comes
from the fact under suitable assumptions, the image of $\Phi$
under the Hecke operator has no pole, and {\it has therefore to be
a constant $\lambda$} \cite{Mayr:1993mq}. The relevant constants
are listed in Appendix B.3. This observation is at the heart of
the simplifications that allow the heterotic-type II duality to
work. In that case, we can thus rewrite \eqref{mi3} as
\begin{equation}
\label{mi4} I_{d,d}[\Phi]= 2^{\frac{d}{2}+1} \lambda \int_{{\cal F}}
\frac{d^2\tau}{\tau_2^2} Z(g/2,b/2;\tau) \ .
\end{equation}
This is now a standard integral $I_d=\int Z_{d,d}(g,b,\tau)$ over
the fundamental domain of $Sl(2,\Zint)$, which can be for instance
represented in terms of Eisenstein series \cite{Obers:1999um}. For
$d=1,2,3,4$, we recall in particular
\begin{subequations}
\label{Iint}
\begin{eqnarray}
\label{eis} I_1(R)&=&\frac{\pi}{3} \left( R+ \frac{l_s^2}{R} \right)\ ,
\label{I1}
\\ I_2(T,U)&=&-\log \frac{8\pi e^{1-\gamma_E}}{3\sqrt{3}} T_2 U_2
|\eta(U)|^4 |\eta(T)|^4 \ , \label{I2} \\
I_3(g,b)&=&\frac{1}{\pi}
\eis{SO(3,3,\Zint)}{4}{s=1}=\frac{1}{\pi}
\eis{Sl(4,\Zint)}{4}{s=1} \ , \label{I3}
\\ I_4(g,b)&=&\frac{1}{\pi}
\eis{SO(4,4,\Zint)}{V}{s=1}= \frac{1}{\pi}
\eis{SO(4,4,\Zint)}{C}{s=1} \ ,  \label{I4}
\end{eqnarray}
\end{subequations}
where the normalization here differs from that of \cite{Obers:1999um}.
For a given discrete duality symmetry group $G(\Zint)$,
the order $s$ Eisenstein series of representation
$\R$ is defined by,
\begin{equation}
\label{geneis}
\eis{G(\Zint)}{\R}{s} = \sum_{m \in \Lambda_\R\backslash\{0\} }
\delta( m\wedge m) \left[ {\M^2 (\R)} \right]^{-s}
\end{equation}
where we refer to \cite{Obers:1999um} for explicit expressions of the
$G(\Zint)$-invariant BPS masses $\M^2(\R)$ and the
half-BPS condition $m \wedge m =0$, that are relevant for the cases in
\eqref{I3}, \eqref{I4}.

The simplification that occurred in the computation of
\eqref{modint} is actually of much more general validity, and would hold
provided the shifted lattice sum can be rewritten as $Z(2\tau)$ for some
modular invariant function $\tau$. The insertion $\Phi$ can then
be replaced by its value, {\it when constant}, under the Hecke
operator $H_{\Gamma_2^-}$:
\begin{equation}
\label{thm} I_{d,d}[\Phi]= \frac{2\lambda}{3} I_{d,d}[1 ] \quad
\mbox{if}\quad H_{\Gamma_2^-}(\Phi)=\lambda \ .
\end{equation}
The same also holds for fractional shifts $1/n$, $n>2$ that occur
in $\Zint_n$ orbifolds, although we will not explore this topic.
We also mention that the rule \eqref{thm} holds as well in the presence
of insertions of momenta $p_R^i, p_L^i$.

\section{Useful modular identities \label{modu} }
We refer to Appendix F of \cite{Kiritsis:1997hj} for generalities
and useful identities on modular forms. Here we list the modular
identities that are useful for the present work.

\subsection{Theta functions and their derivatives \label{moduth}}
Our conventions for the Jacobi Theta functions are
\begin{equation}
\label{t0} \th\ar{a}{b}(v,\tau)=\sum_{n\in \Zint} q^{\frac12
\left(n-\frac{a}{2}\right)^2} e^{\left(v-i\pi b\right)
\left(n-\frac{a}{2}\right)}\ ,\quad q=e^{2\pi i \tau} \ ,
\end{equation}
where the normalization of $v$ is non-standard. We also use the
Erderlyi notation
\begin{equation}
\label{t01} \th_1=\th\ar{1}{1}\ ,\ \th_2=\th\ar{1}{0}\ ,\
\th_3=\th\ar{0}{0}\ ,\ \th_4=\th\ar{0}{1} \ .
\end{equation}
Note that $\th_{2,3,4}$ are even functions of their argument $v$
while $\th_1$ is odd, and $\th_1'=-i \eta^3$ where we denote by a
prime the differentiation with respect to $v$. For more than one
derivation $\partial/\partial v$, the result is not modular
covariant anymore, and has to be corrected by non-holomorphic
contributions, analogous to the replacement $E_2\to
\et=E_2-3/(\pi\tau_2)$. We use the multiprime symbols for the
result of this covariantization. We hence have
\begin{subequations}
\begin{equation}
\label{t1}
 \frac{\th_2^{''}}{\th_2}=
\frac{1}{12}\left( \et+\th_3^4+\th_4^4\right)
\end{equation}
\begin{equation}
\label{t2}
 \frac{\th_3^{''}}{\th_3}=
\frac{1}{12}\left( \et+\th_2^4-\th_4^4\right)
\end{equation}
\begin{equation}
\label{t3}
 \frac{\th_4^{''}}{\th_4}=
\frac{1}{12}\left( \et-\th_2^4-\th_3^4\right)
\end{equation}
\begin{equation}
\label{t4}
 \frac{\th_2^{''''}}{\th_2}=
\frac{1}{48}\left(-2
  E_4+\et^2+2\et(\th_3^4+\th_4^4)+3\th_2^8\right)
\end{equation}
\begin{equation}
\label{t5}
 \frac{\th_3^{''''}}{\th_3}=
\frac{1}{48}\left(-2
  E_4+\et^2+2\et(\th_2^4-\th_4^4)+3\th_3^8\right)
\end{equation}
\begin{equation} \frac{\th_4^{''''}}{\th_4}=
\label{t6} \frac{1}{48}\left(-2
  E_4+\et^2-2\et(\th_2^4+\th_3^4)+3\th_4^8\right) \ .
\end{equation}
\end{subequations}
The following combinations will be particularly relevant
\begin{subequations}
\begin{equation}
\label{t7} \frac{\th_3^{''''}}{\th_3} +\frac{\th_4^{''''}}{\th_4}
- 3
 \left( \frac{\th_3^{''}}{\th_3} \right)^2 -3
 \left( \frac{\th_4^{''}}{\th_4} \right)^2
=-\frac18 \th_2^8
\end{equation}
\begin{equation}
\label{t10} \left(\frac{\th_3^{''}}{\th_3}\right)^2 +
\left(\frac{\th_4^{''}}{\th_4}\right)^2 -2
\frac{\th_3^{''}}{\th_3}   \frac{\th_4^{''}}{\th_4} =\frac1{16}
\th_2^8
\end{equation}
\end{subequations}
and the following identities are useful to make contact with
\cite{Lerche:1998nx}:
\begin{subequations}
\begin{equation}
\label{t8} \left(\frac{\th_3^{''}}{\th_3}\right)^2
 +\left(\frac{\th_4^{''}}{\th_4}\right)^2
=\frac1{72}
\left(\et-\frac{\th_3^4+\th_4^4}{2}\right)^2+\frac1{32} \th_2^8
\end{equation}
\begin{equation}
\label{t9} 2\frac{\th_3^{''}}{\th_3} \frac{\th_4^{''}}{\th_4}
=\frac1{72}
\left(\et-\frac{\th_3^4+\th_4^4}{2}\right)^2-\frac1{32} \th_2^8 \ .
\end{equation}
\end{subequations}

\subsection{Summation identities \label{modusum}}
In our ``modular Einstein convention'', $\alpha=2,3,4$ is summed
over all even spin structures.
The following equations are useful to convert the contribution of
the unshifted orbit into a sum of shifted orbits:
\begin{subequations}
\begin{equation}
\label{s1} \th_\alpha^{16} - \left[ 2{\th_3^8\th_4^8} \orb
\right]=0
\end{equation}
\begin{equation}
\label{s2} \th_\alpha''~\th_\alpha^{15} - \left[
{\th_3^8\th_4^8}\left( \frac{\th_3^{''}}{\th_3}
+\frac{\th_4^{''}}{\th_4} \right) \orb \right]=0
\end{equation}
\begin{equation}
\label{s3} \th_\alpha''''~\th_\alpha^{15} -\left[
{\th_3^8\th_4^8}\left( \frac{\th_3^{''''}}{\th_3}
+\frac{\th_4^{''''}}{\th_4} \right) \orb \right] = 96 \eta^{24}
\end{equation}
\begin{equation}
\label{s4} (\th_\alpha'')^2~\th_\alpha^{14} -\left[
2{\th_3^8\th_4^8} \frac{\th_3^{''}}{\th_3}
\frac{\th_4^{''}}{\th_4} \orb \right] = 16 \eta^{24}
\end{equation}
\begin{equation}
\label{s5} (\th_\alpha'')^2~\th_\alpha^{14} - \left[
\th_3^8\th_4^8 \left( \left(\frac{\th_3^{''}}{\th_3} \right)^2 +
\left(\frac{\th_4^{''}}{\th_4} \right)^2\right) \orb \right] = -32
\eta^{24}\ .
\end{equation}
\end{subequations}
where $\orb$ denotes the two extra terms obtained from the first
by applying $S$ and $ST$ modular transformations.

\subsection{Hecke identities \label{moduh}}
As proven in Appendix \ref{latint}, insertions of almost holomorphic
modular forms into integrals of projected lattice sums can be
replaced by two-thirds their value $\lambda$ under the Hecke
operator \eqref{hg2-}.
Here we list the corresponding value for the
modular forms of interest
\begin{subequations}
\begin{equation}
\label{h1} H_{\Gamma_2^-} \left[ {\th_3^8\th_4^8}/{\eta^{24}}
\right] =0\ ,\quad H_{\Gamma_2^-} \left[ {\th_3^8\th_4^8}\et
/{\eta^{24}} \right] =0
\end{equation}
\begin{equation}
\label{h2} H_{\Gamma_2^-} \left[ {\th_3^8\th_4^8}
E_4/{\eta^{24}}\right] =360\ ,\quad H_{\Gamma_2^-} \left[
{\th_3^8\th_4^8}\et^2/{\eta^{24}} \right] =72
\end{equation}
\begin{equation}
\label{h3} H_{\Gamma_2^-} \left[ \frac{\th_3^8\th_4^8}{\eta^{24}}
 \left( \frac{\th_3^{''}}{\th_3} +\frac{\th_4^{''}}{\th_4}
\right) \right] =0
\end{equation}
\begin{equation}
\label{h4} H_{\Gamma_2^-} \left[ \frac{\th_3^8\th_4^8}{\eta^{24}}
 \left( \frac{\th_3^{''}}{\th_3} +\frac{\th_4^{''}}{\th_4} \right) \et
 \right]
=24
\end{equation}
\begin{equation}
\label{h5} H_{\Gamma_2^-} \left[ \frac{\th_3^8\th_4^8}{\eta^{24}}
 \left( \frac{\th_3^{''''}}{\th_3} +\frac{\th_4^{''''}}{\th_4} \right)
 \right]
=0
\end{equation}
\begin{equation}
\label{h6} H_{\Gamma_2^-} \left[ \frac{\th_3^8\th_4^8}{\eta^{24}}
 \left( \left(\frac{\th_3^{''}}{\th_3}\right)^2 +
\left(\frac{\th_4^{''}}{\th_4}\right)^2 \right) \right] =16
\end{equation}
\begin{equation}
\label{h7} H_{\Gamma_2^-} \left[ \frac{\th_3^8\th_4^8}{\eta^{24}}
 \left(\frac{\th_3^{''}}{\th_3}\right)
\left(\frac{\th_4^{''}}{\th_4}\right)  \right] =-4 \ .
\end{equation}
\end{subequations}
These formulae can be obtained by looking at the leading $q$ expansions,
or by using the following duplication identities:
\begin{subequations}
\begin{equation}
\th_2(2\tau)=\frac{1}{\sqrt{2}}\sqrt{\th_3^2(\tau)-\th_4^2(\tau)}\ ,\quad
\th_3(2\tau)=\frac{1}{\sqrt{2}}\sqrt{\th_3^2(\tau)+\th_4^2(\tau)}
\end{equation}
\begin{equation}
\th_4(2\tau)=\sqrt{\th_3(\tau)\th_4(\tau)}
\ ,\quad
\eta(2\tau)=2^{-2/3}\th_2^{2/3}(\tau)(\th_3(\tau)\th_4(\tau))^{1/6}
\end{equation}
\begin{equation}
\th_2(\tau/2)=\sqrt{2\th_2(\tau)\th_3(\tau)}\ ,\quad
\th_3(\tau/2)=\sqrt{\th_3^2(\tau)+\th_2^2(\tau)}
\end{equation}
\begin{equation}
\th_4(\tau/2)=\sqrt{\th_3^2(\tau)-\th_2^2(\tau)}\ ,\quad
\eta(\tau/2)=2^{-1/6}\th_4^{2/3}(\tau)(\th_2(\tau)\th_3(\tau))^{1/6}
\end{equation}
\begin{equation}
\th_2\left(\frac{\tau+1}{2}\right)=e^{\frac{i\pi}{ 8}}
\sqrt{2\th_2(\tau)\th_4(\tau)}\ ,\quad
\th_3\left(\frac{\tau+1}{2}\right)=\sqrt{\th_4^2(\tau)+i\th_2^2(\tau)}
\end{equation}
\begin{equation}
\th_4\left(\frac{\tau+1}{2}\right)=\sqrt{\th_4^2(\tau)-i\th_2^2(\tau)}\ ,\quad
\eta\left( \frac{\tau+1}{2}\right)=2^{-1/6}~e^{\frac{i\pi}{24}}~\th_3^{2/3}(\tau)(\th_2(\tau)\th_4(\tau))^{1/6}
\end{equation}
\begin{equation}
\th_2(\tau)=2\frac{\eta^2(2\tau)}{\eta(\tau)}\ ,\quad
\th_4(\tau)=\frac{\eta^2(\tau/2)}{\eta(\tau)}\ ,\quad
\th_3(\tau)=e^{i \pi/12}\frac{\eta^2((\tau+1)/2)}{\eta(\tau)}
\end{equation}
\begin{equation}
\eta(2\tau)~
\eta(\tau/2)~
\eta\left( (\tau+1)/2 \right)
=e^{-i \pi/24} \eta^3(\tau) \ .
\end{equation}
\end{subequations}

\providecommand{\href}[2]{#2}\begingroup\raggedright
\endgroup

\end{document}